\documentclass[12pt]{report}
\usepackage{amsmath}
\usepackage{latexsym}
\usepackage{amsfonts}
\usepackage[normalem]{ulem}
\usepackage{soul}
\usepackage{array}
\usepackage{amssymb}
\usepackage{extarrows}
\usepackage{graphicx}
\usepackage[backend=biber,
style=numeric,
sorting=none,
isbn=false,
doi=false,
url=false,
]{biblatex}

\usepackage{subfig}
\usepackage{wrapfig}
\usepackage{wasysym}
\usepackage{enumitem}
\usepackage{adjustbox}
\usepackage{ragged2e}
\usepackage[svgnames,table]{xcolor}
\usepackage{tikz}
\usepackage{longtable}
\usepackage{changepage}
\usepackage{setspace}
\usepackage{hhline}
\usepackage{multicol}
\usepackage{tabto}
\usepackage{float}
\usepackage{multirow}
\usepackage{makecell}
\usepackage{fancyhdr}
\usepackage[toc,page]{appendix}
\usepackage[hidelinks]{hyperref}
\usetikzlibrary{shapes.symbols,shapes.geometric,shadows,arrows.meta}
\tikzset{>={Latex[width=1.5mm,length=2mm]}}
\usepackage{flowchart}\usepackage[paperheight=11.69in,paperwidth=8.27in,left=1.0in,right=1.0in,top=1.0in,bottom=1.0in,headheight=1in]{geometry}
\usepackage[utf8]{inputenc}
\usepackage[T1]{fontenc}
\TabPositions{0.5in,1.0in,1.5in,2.0in,2.5in,3.0in,3.5in,4.0in,4.5in,5.0in,5.5in,6.0in,}

\renewcommand{\_}{\kern-1.5pt\textunderscore\kern-1.5pt}

\setlistdepth{9}
\renewlist{enumerate}{enumerate}{9}
		\setlist[enumerate,1]{label=\arabic*)}
		\setlist[enumerate,2]{label=\alph*)}
		\setlist[enumerate,3]{label=(\roman*)}
		\setlist[enumerate,4]{label=(\arabic*)}
		\setlist[enumerate,5]{label=(\Alph*)}
		\setlist[enumerate,6]{label=(\Roman*)}
		\setlist[enumerate,7]{label=\arabic*}
		\setlist[enumerate,8]{label=\alph*}
		\setlist[enumerate,9]{label=\roman*}

\renewlist{itemize}{itemize}{9}
		\setlist[itemize]{label=$\cdot$}
		\setlist[itemize,1]{label=\textbullet}
		\setlist[itemize,2]{label=$\circ$}
		\setlist[itemize,3]{label=$\ast$}
		\setlist[itemize,4]{label=$\dagger$}
		\setlist[itemize,5]{label=$\triangleright$}
		\setlist[itemize,6]{label=$\bigstar$}
		\setlist[itemize,7]{label=$\blacklozenge$}
		\setlist[itemize,8]{label=$\prime$}

\title{Putative cell type discovery from single-cell gene expression data}
\date{}

\begin{document}

\maketitle
\par

\vspace{\baselineskip}
\begin{justify}
Zhichao Miao\textsuperscript{1,2}, Pablo Moreno\textsuperscript{1}, Ni Huang\textsuperscript{1,2}, Irene Papatheodorou\textsuperscript{1}, Alvis Brazma\textsuperscript{1}$\ast$ , Sarah A Teichmann\textsuperscript{2,3}$\ast$ 
\end{justify}\par

\vspace{\baselineskip}
\begin{enumerate}
	\item European Molecular Biology Laboratory, European Bioinformatics Institute (EMBL-EBI), Wellcome Genome Campus, Cambridge, CB10 1SD, UK\par

	\item Wellcome Sanger Institute, Wellcome Genome Campus, Hinxton, Cambridge, CB10 1SA, UK\par

	\item Department of Physics, Cavendish Laboratory, JJ Thompson Ave, University of Cambridge, Cambridge, CB3 0EH, UK
\end{enumerate}\par

\vspace{\baselineskip}
{\fontsize{8pt}{9.6pt}\selectfont $\ast$ To whom correspondence should be addressed\par}\par

\href{mailto:st9@sanger.ac.uk}{\ul{st9@sanger.ac.uk}}\par

\href{mailto:brazma@ebi.ac.uk}{\ul{brazma@ebi.ac.uk}}\par

\subsection*{Abstract}
\addcontentsline{toc}{subsection}{Abstract}

\vspace{\baselineskip}
\begin{justify}
We present a novel method for automated identification of putative cell types from single-cell RNA-seq (scRNA-seq) data. By iteratively applying a machine learning approach to an initial clustering of gene expression profiles of a given set of cells, we simultaneously identify distinct cell groups and a weighted list of \textit{feature} \textit{genes} for each group. The feature genes, which are differentially expressed in the particular cell group, jointly discriminate the given cell group from other cells. Each such group of cells corresponds to a putative cell type or state, characterised by the feature genes as markers. To benchmark this approach, we use expert-annotated scRNA-seq datasets from a range of experiments, as well as comparing to existing cell annotation methods, which are all based on a pre-existing reference. We show that our method automatically identifies the $``$ground truth$"$  cell assignments with high accuracy. Moreover, our method, $``$Single Cell Clustering Assessment Framework$"$  (SCCAF) predicts new putative biologically meaningful cell-states in published data on haematopoiesis and the human cortex. SCCAF is available as an open-source software package on GitHub (\href{https://github.com/SCCAF/sccaf}{\ul{https://github.com/SCCAF/sccaf}}) and as a Python package index and has also been implemented as a Galaxy tool in the Human Cell Atlas. 
\end{justify}\par

\vspace{\baselineskip}
\subsection*{Introduction}
\addcontentsline{toc}{subsection}{Introduction}
\subsection*{Identifying cell types in multicellular organisms and understanding the relationships between them has been a major aim of biological research since the discovery of cells by Robert Hooke almost 400 years ago. Historically, cell types have been defined by their morphology as assessed by microscopy, by their locations in an organism, by their function in vivo or in vitro, or by their developmental and evolutionary history . In immunology, the advent of flow cytometry led to an assignment of cell types according to a small number of molecular markers on the cell surface. With the advent of high-throughput molecular techniques, a subtler classification of cells by their molecular profiles has become possible. Data from single-cell RNA-seq (scRNA-seq) is one of the most data-rich and high-dimensional sources of information for the discovery of new putative cell types and refining the classification of existing ones. Defining and identifying all cell types in a human body is one of the goals of the Human Cell Atlas (HCA) project, which aims to apply scRNA-seq to a representative sample of all human cells . }
\addcontentsline{toc}{subsection}{Identifying cell types in multicellular organisms and understanding the relationships between them has been a major aim of biological research since the discovery of cells by Robert Hooke almost 400 years ago. Historically, cell types have been defined by their morphology as assessed by microscopy, by their locations in an organism, by their function in vivo or in vitro, or by their developmental and evolutionary history . In immunology, the advent of flow cytometry led to an assignment of cell types according to a small number of molecular markers on the cell surface. With the advent of high-throughput molecular techniques, a subtler classification of cells by their molecular profiles has become possible. Data from single-cell RNA-seq (scRNA-seq) is one of the most data-rich and high-dimensional sources of information for the discovery of new putative cell types and refining the classification of existing ones. Defining and identifying all cell types in a human body is one of the goals of the Human Cell Atlas (HCA) project, which aims to apply scRNA-seq to a representative sample of all human cells . }
\setlength{\parskip}{9.96pt}
\begin{justify}
A typical way of using scRNA-seq data for putative cell type identification relies on clustering the cells in a sample by their expression profiles, followed by an expert inspection of each cluster, for instance, looking for known cell marker genes, the pathways expressed in the cluster, or selective comparison of cells within and between clusters. To assist human experts, computational analysis tools, such as SCANPY\href{https://paperpile.com/c/BKye4I/VJJ9r}{\textsuperscript{8}} and Seurat\href{https://paperpile.com/c/BKye4I/cthgZ}{\textsuperscript{9}} cluster the cells by different methods\href{https://paperpile.com/c/BKye4I/BggCD+lCRQo+Jazbe}{\textsuperscript{10–12}} and then visualize the clustering using dimensionality reduction methods. To date, a large number of scRNA-seq datasets have been annotated this way, from mouse, human and other organisms, and made publicly available \href{https://paperpile.com/c/BKye4I/d4ibg+vAx8W+dzVJ6+Ve8gd+ovLNj+1T9D1+3ZMeu}{\textsuperscript{13–19}}. To some extent, such expert annotated datasets can be considered as ground-truth reflecting the current state of knowledge in the field. 
\end{justify}\par

\begin{justify}
Although at the current state of play involvement of human experts in annotating scRNA-seq datasets is inevitable, approaches based on clustering and human inspection suffer from significant drawbacks. First, typically, it is not known \textit{a priori }how many different classes of cells the studied sample contains. Methods to assess the likely number of clusters in data have been developed \href{https://paperpile.com/c/BKye4I/4AUxR}{\textsuperscript{20}} and distance-based cluster merging has been proposed to optimize clustering\href{https://paperpile.com/c/BKye4I/NRIXy}{\textsuperscript{21}}, nevertheless, even these methods cannot guarantee that the resulting clusters represent biologically distinct groups of cells corresponding to putative cell states or types. Second, with growing amounts of data, a careful human expert inspection of each dataset is not scalable. For instance, the HCA project will be generating data at scale, where the human inspection will become a bottleneck. To facilitate annotation of new datasets, various automated methods that exploit previously identified cell types and knowledge of their markers have been proposed \href{https://paperpile.com/c/BKye4I/z6JA+ia4B+mL7w+cWTh+YStx+1lv9+J2VQ+zaRi+kYlT}{\textsuperscript{22–30}}. These \textit{reference-based methods,} however, cannot discover new, yet unannotated types of cells or new cell states. 
\end{justify}\par

\begin{justify}
Here we propose an automated method that allows for the potential discovery of novel, not yet annotated putative cell types. (Here we treat the terms $``$putative cell type$"$  and $``$cell state$"$  loosely as synonyms referring to a biologically meaningful group of cells characterised by a specific and well-defined pattern of gene expression). Our method, which we call Single Cell Clustering Assessment Framework (SCCAF), is based on the iterative application of machine learning and self-projection\href{https://paperpile.com/c/BKye4I/Ve8gd}{\textsuperscript{16}} to clusters, thus gradually merging the clusters that correspond to the same putative cell type. and assigning a set of defining feature genes to each group of cells. 
\end{justify}\par

\begin{justify}
The validity of our approach is based on three assumptions. First, like in any scRNA-seq approach, we assume that each putative cell type is defined by a specific RNA expression profile. Second, for benchmarking purposes, we are assuming here that the published, human expert annotated scRNA-seq datasets represent the ground truth. Third, we assume that if an automated method accurately reconstructs human annotation in diverse scRNA-seq datasets, it is also likely to work well on new datasets. We test these assumptions on simulated as well as real scRNA-seq datasets from a range of experiments, and empirically conclude that our method is able to reconstruct the cell types as annotated by human experts with high accuracy. Moreover, we use a $``$transparent-box$"$  classifier building a computational model defining the respective cell group \textit{via} a weighted list of feature genes for each cluster. We found that our feature genes often include the known marker genes for characterised cell types. 
\end{justify}\par

\begin{justify}
With the throughput of scRNA-seq experiments increasing, and potentially billions of cells being profiled in projects like the Human Cell Atlas (HCA) initiative\href{https://paperpile.com/c/BKye4I/rvY0t}{\textsuperscript{7}}, automated cell type annotation will become increasingly important. As already noted, unlike the previously published cell-type annotation methods, which use reference annotations, we aim to automatically identify biologically meaningful groups of cells that can be discriminated from all other cells \textit{via} models based purely on the expression of appropriate feature genes. Our approach can be combined with reference-based methods, first to identify meaningful clusters, and then to annotate them, e.g. to assign standardised names derived from earlier datasets. Our method is implemented as an open-source software available at \href{https://github.com/SCCAF/sccaf}{\ul{https://github.com/SCCAF/sccaf}}. It has been developed as a part of our implementation of a high-throughput data analysis pipeline that allows for unified automated processing of all publicly available scRNA-seq data, including the data generated by the HCA project.
\end{justify}\par

\subsection*{Results}
\addcontentsline{toc}{subsection}{Results}
\subsubsection*{A self-projection-based approach}
\addcontentsline{toc}{subsubsection}{A self-projection-based approach}
\begin{justify}
The input to our method is a matrix of real values, each column representing a cell and each row, representing a gene (\textbf{Figure 1a}). More specifically, the input is a cell by gene matrix of gene expression values as measured in a scRNA-seq experiment. First, a clustering algorithm (e.g., k-means, Louvain\href{https://paperpile.com/c/BKye4I/Jazbe}{\textsuperscript{12}} or Leiden\href{https://paperpile.com/c/BKye4I/BggCD}{\textsuperscript{10}} clustering) is applied to the columns (\textit{i.e.} to the cells); the clustering can be based either on the entire set of genes, the highly variable genes or done in Principal Component (PC) space using a chosen number of PCs. Then, each cluster is split into a training and a test set, and a classifier is trained and then applied to the test set to measure how well the model discriminates between the cells in the particular cluster over all other cells in the test set. Comparing the predicted clusters to the original ones is known as \textit{self-projection\href{https://paperpile.com/c/BKye4I/Ve8gd}{\textsuperscript{16}}}. The \textit{self-projection accuracy} is defined as the percentage of correctly predicted cells in the entire dataset and can be used to assess the reliability of the clustering. The comparison between the predicted cluster and the actual clusters in the test set gives us the \textit{confusion rate }for each cluster and the respective\textit{ confusion matrix} (see \textbf{Methods}). If the data is $``$over-clustered$"$ , \textit{i.e.}, if two or more clusters, in fact, represent the same type of cells, the classifier will not be able to discriminate between these clusters, and the respective confusion rate will be high. Detecting $``$under-clustering$"$  is more challenging, but as our computational experiments show, if a cluster represents a mixture of cell types, the classifier performance will typically be poorer than for a cluster representing a well-defined group of cells (\textbf{Suppl Figure S1}). 
\end{justify}\par

\vspace{\baselineskip}
\begin{justify}
Next, we normalise the confusion matrix (\textbf{Figure 2}) by computing the ratio of the misclassified cells over the correctly assigned cells to account for clusters of different sizes and then discretise it by applying a maximum confusion threshold derived from the maximum confusion rate of the entire dataset (see \textbf{Methods}). The discretized confusion matrix represents a cluster connection graph (intuitively describing the similarity between clusters); the connected clusters are merged according to this graph. We iterate the machine learning and self-projection approach, merging the connected clusters until the overall \textit{self-projection accuracy} keeps growing, or reaches 98$\%$  default cutoff. 
\end{justify}\par

\vspace{\baselineskip}
\begin{justify}
Our assumption is that if there is a set of genes expressed differentially in a cluster in comparison to other cells in the training set, such \textit{feature genes} can be discovered by machine learning. Thus, applying the trained model to the test data, the clusters representing putative cell types will be restored reasonably well. If, on the contrary, a model derived from the training half of the dataset is $``$confused$"$  (unable to find $``$good$"$  discriminative genes), this indicates that there is no good set of feature genes whose expression define this cluster unambiguously. Thus, in the context of expression data, a cluster without discriminative genes cannot be considered as a biologically coherent and distinct group of cells. Specifically, in our approach, we use logistic regression and 5-fold cross-validation (hence, by default we require a minimum of 10 cells in a cluster), which was shown by Ntranos \textit{et al.\href{https://paperpile.com/c/BKye4I/var3u}{\textsuperscript{31}}} as a fast and effective method for capturing differentially expressed genes. Four other machine learning models (Support Vector Machine, Decision Tree, Random Forest and Gaussian Naive Bayes) are also implemented in the SCCAF framework and have been tested by us (\textbf{Figure S2}). In the tests on both simulated data and real data (\textbf{Figure 1} and \textbf{Figure S3}), the self-projection results indicate that confusion of cell cluster assignments happens almost exclusively between cell clusters of the same cell type, rather than between clusters of different cell types. 
\end{justify}\par

\subsubsection*{Evaluating SCCAF on simulated data}
\addcontentsline{toc}{subsubsection}{Evaluating SCCAF on simulated data}
\begin{justify}
To test our method, we used both simulated and expert-annotated datasets. We used two simulation approaches: first, based on the multivariable normal distribution (see \textbf{Methods}) and second based on the scRNA-seq data simulator called Splatter\href{https://paperpile.com/c/BKye4I/fFIer}{\textsuperscript{32}}. 
\end{justify}\par

\vspace{\baselineskip}
\begin{justify}
We first simulated 3000 cells of 6 cell types using a multivariable normal distribution. Each cell type includes 500 cells of 10-20 feature genes, \textbf{Figure 2a}. A logistic regression model was trained on half of the dataset and applied to the whole dataset. The original cell clusters and the projection results are plotted in \textbf{Figure 2a }as t-SNE\href{https://paperpile.com/c/BKye4I/Tsuir}{\textsuperscript{33}}. The projection result is identical to the original assignment, as the cell type-related features are captured in the logistic regression model, and the self-projection assessment accuracy is above 94$\%$ . 
\end{justify}\par

\vspace{\baselineskip}
\begin{justify}
As logistic regression weighs each gene in a linear model, we extracted the top-weighted genes and compared them with the feature genes used in the simulation (\textbf{Figure 2f)}. The top-weighted genes recover >75$\%$  of the simulated feature genes, and both lists overlap well. While this result is not surprising, as our simulation method and a logistic regression model are well-matched, the multivariate simulation approach has been previously\href{https://paperpile.com/c/BKye4I/var3u}{\textsuperscript{31}} demonstrated to capture important features of scRNA-seq data. 
\end{justify}\par

\vspace{\baselineskip}
\begin{justify}
We also simulated cell populations where only combinations of features could discriminate between the cell types unambiguously (for instance, the cell types \textit{A} and \textit{B} are jointly discriminated from the third cell type \textit{C} by expression of gene \textit{a}, while the cell type \textit{A} is discriminated from the cell type \textit{B} by the expression of gene \textit{b}). Using the same SCCAF approach in the PC-space, we find the optimal cell clustering can be achieved, shown in \textbf{Figure S4}. 
\end{justify}\par

\vspace{\baselineskip}
\begin{justify}
We reach a similar conclusion using data generated by Splatter\href{https://paperpile.com/c/BKye4I/fFIer}{\textsuperscript{32}}, as shown in \textbf{Figure S5}. We simulated six cell types of different numbers ranging from 252 to 1072 cells, also using a hierarchy of simulations with different similarities between cell clusters. The \textit{de.prob} parameter in Splatter indicates the differential expression probability between cell clusters, higher values indicate higher differential expression levels. In \textbf{Figure S6}, we simulate a six cell-type dataset with de.prob range from 0.4 to 0.8. The results show that the Louvain clustering resolution 1 includes over clustering when de.prob is above 0.5. However, SCCAF finds clusters identical to the original cell type assignments. 
\end{justify}\par

\vspace{\baselineskip}
\subsubsection*{Testing the method on expert-annotated scRNA-seq datasets }
\addcontentsline{toc}{subsubsection}{Testing the method on expert-annotated scRNA-seq datasets }
\begin{justify}
To assess to what extent SCCAF recovers expert-annotated groups of cells, we tested the method on 8 published and annotated scRNAseq data sets. First, we used the mouse retina data set from Shekhar \textit{et al.} \href{https://paperpile.com/c/BKye4I/d4ibg}{\textsuperscript{13}}. For this dataset, the outputs from Louvain clusterings in the PC space at resolution 1.0 (\textbf{Figure 3a)} show over-clustering of the Rod Bipolar cells, while resolution 0.3 clustering shows both under-clustering and over-clustering in comparison to the expert annotation. SCCAF starts from Louvain clustering resolution 1.0 and goes through two rounds of optimizations (\textbf{Figure S7}). The cluster connection graph in the first round of the SCCAF optimization, \textbf{Figure 3b}, shows that the clusters containing cells of the same cell-type (as described in the publication) are highly confused. By merging the clusters SCCAF gradually reduces the normalized confusion rate until the total self-projection accuracy exceeds the chosen cutoff of 95$\%$ . The accuracies of the cross-validation and on the test set continuously increase during the two rounds of optimization from 74$\%$  to 98$\%$ , \textbf{Figure 3c}. As the t-SNE plots in \textbf{Figure 3d }shows, the SCCAF confusion matrix-based cluster optimization achieves a similar result to the ‘ground truth’ published cell annotation. The river plot comparing SCCAF-optimized clustering and the expert annotated dataset in \textbf{Figure 3e} shows that they are almost perfectly matched, the adjusted rand index\href{https://paperpile.com/c/BKye4I/cbqJ}{\textsuperscript{34}} is over 0.99. Thus, in this dataset, the $``$ground truth$"$  can be recovered almost automatically with high accuracy. Furthermore, the feature genes that discriminate the cell clusters may facilitate any subsequent manual annotation of the clusters. 
\end{justify}\par

\vspace{\baselineskip}
\begin{justify}
To test our method more broadly, we used 7 additional published datasets annotated by experts, assuming the annotated cell types represent the $``$ground truth$"$ . Specifically, we used human and mouse pancreas datasets from Baron \textit{et al.\href{https://paperpile.com/c/BKye4I/vAx8W}{\textsuperscript{14}}}, a mouse cortex dataset from Zeisel \textit{et al.\href{https://paperpile.com/c/BKye4I/Gfzwj}{\textsuperscript{35}}}, a retinal bipolar neuron dataset from Shekhar \textit{et al.\href{https://paperpile.com/c/BKye4I/d4ibg}{\textsuperscript{13}}}, a Smart-Seq2 human pancreatic islets dataset from Segerstolpe \textit{et al.\href{https://paperpile.com/c/BKye4I/mJd6E}{\textsuperscript{36}}}, a inDrops mouse visual cortex dataset from Hrvatin \textit{et al.\href{https://paperpile.com/c/BKye4I/eE5EA}{\textsuperscript{37}}}, a mCEL-seq2 human liver dataset from Aizarani \textit{et al.\href{https://paperpile.com/c/BKye4I/otnx}{\textsuperscript{38}}} and a SMART-Seq mouse cortex dataset from Tasic \textit{et al.\href{https://paperpile.com/c/BKye4I/AXkh}{\textsuperscript{39}}}. Cell type annotations were obtained from the original publications. In almost all cases, SCCAF achieves clusterings close to the manual annotation of published clusters (\textbf{Figure S8 }and\textbf{ Figure S9})\textbf{ }average adjusted rand index >0.94. 
\end{justify}\par

\vspace{\baselineskip}
\begin{justify}
A key challenge in scRNA-seq data analysis is finding the correct number of cell clusters in a data set. For instance, SC3\href{https://paperpile.com/c/BKye4I/4AUxR}{\textsuperscript{20}} clustering attempts to estimate the optimal number of cell clusters using Tracy-Widom theory\href{https://paperpile.com/c/BKye4I/CeZ6x}{\textsuperscript{40}} on random matrices. In \textbf{Figure 4}, the self-projection clustering optimization process starts from a Louvain clustering of resolution 3.5, in which many cells are over-clustered. The self-projection clustering optimization stops after four rounds when the self-projection accuracy is > 96$\%$ . We further lower the confusion rate threshold until the results are under-clustered (\textbf{Figure 4b}). For all the clustering $``$snapshots$"$  throughout nine rounds of optimisation, self-projection clustering assessment was repeated 100 times by random sampling the training set and the test set. According to the distribution of the self-projection accuracy, the optimal clustering shows better accuracy than over-clustering or under-clustering. In addition, the standard deviation of the self-projection accuracy upon random sampling is smaller for the optimal clustering than other cluster solutions (\textit{i.e} over-clustering or under-clustering, or intermediate optimisation rounds). We also tested our method on a SMART-seq2 dataset, specifically the pancreas islet cell dataset \href{https://paperpile.com/c/BKye4I/dzVJ6}{\textsuperscript{15}}, which revealed similar results (\textbf{Figure 4c}, \textbf{d)}. Our computational experiments show (\textbf{Figure S1}) that in real-world datasets, self-projection accuracy is the highest for the cell-type assignments corresponding to the ground-truth in almost all cases. (In rare cases it may produce under-clustering.) Therefore, in most cases, the iterations can run until the self-projection accuracy stops increasing, while in some rare cases user intervention is needed. 
\end{justify}\par

\subsubsection*{SCCAF defines cell states in erythroid maturation}
\addcontentsline{toc}{subsubsection}{SCCAF defines cell states in erythroid maturation}
\begin{justify}
A recent mouse haematopoietic stem cell (HSCs) differentiation dataset from Tusi \textit{et al.} \href{https://paperpile.com/c/BKye4I/RdCh9}{\textsuperscript{41}} characterizes cell states in a continuous differentiation process. Applying SCCAF to a dataset describing a continuous process will lead to clusters that correspond to the most populated regions in the differentiation trajectory. We sought to define these major cell states during the differentiation, by starting from a Louvain clustering at resolution 1.5 (\textbf{Figure 5a}). This results in an optimized clustering corresponding to 12 putative cell states (\textbf{Figure 5b}), with a self-projection assessment accuracy of 92$\%$  on the test data. Several cell clusters are closely related to their annotation reported in the literature (\textbf{Figure S10)}. The logistic regression model derived from SCCAF shows a good discrimination of all the cell clusters (\textbf{Figure 5c}). The top-weighted feature genes in the logistic regression model are retrieved and 35 of them were reported as marker genes in the publication\href{https://paperpile.com/c/BKye4I/RdCh9}{\textsuperscript{41}} and used as probes in the RT-PCR validation (\textbf{Figure 5d}). Thus, the logistic regression weights do correlate to the biological meaning of the cell clusters. 
\end{justify}\par

\vspace{\baselineskip}
\begin{justify}
The erythrocytes further cluster into three cell states: committed erythroid progenitors (CEP), also described in the literature, and early and late erythroid terminal differentiated cells (ETDe and ETDl), which we propose as novel distinct states of erythroid maturation. These three cell states show clear separation in the PC-space (\textbf{Figure 5e}). The three cell states (1,7,6) can be clearly separated by the first two components, which implies that these cell states can be mathematically discriminated based on their transcriptomes. According to differential expression analysis, \textbf{Figure S11}, both upregulated and downregulated genes can be detected between these states, providing further confidence in their biological relevance. 
\end{justify}\par

\vspace{\baselineskip}
\begin{justify}
Finally, we projected the logistic regression models trained on data from Tusi \textit{et al.\href{https://paperpile.com/c/BKye4I/RdCh9}{\textsuperscript{41}}} to an independent mouse hematopoiesis dataset from Giladi \textit{et al.\href{https://paperpile.com/c/BKye4I/DmgXV}{}\textsuperscript{42}}. The majority of the cell populations are re-capitulated, and a UMAP\href{https://paperpile.com/c/BKye4I/cgeDc+fFdsG}{\textsuperscript{43,44}} plot shows an identical distribution (\textbf{Figure 5f}). The erythroid and granulocytic neutrophil branches show the same order of cell clusters (2->4->1->7->6 and 3->0->5). Focusing on the three erythroid states in the Giladi dataset, the self-projection accuracy is 90$\%$ , indicating little confusion between the states. 
\end{justify}\par

\vspace{\baselineskip}
\begin{justify}
Tusi \textit{et al.\href{https://paperpile.com/c/BKye4I/RdCh9}{}\textsuperscript{41}} defined transcriptional events of the erythroid trajectory: they propose Gata1, Epor and Tfrc as induced early in the trajectory. They defined the stages of CEP and ETD, but the did not characterise the early versus late stages in ETD. By looking into the two mouse hematopoiesis datasets from Tusi \textit{et al.\href{https://paperpile.com/c/BKye4I/RdCh9}{\textsuperscript{41}}} and Giladi \textit{et al.\href{https://paperpile.com/c/BKye4I/DmgXV}{\textsuperscript{42}}}\textbf{ }(\textbf{Figure 5g}), we found the three cell stages in erythroid maturation can be discriminated using the logistic regression model. The feature genes encoded in the CEP stage detects the known markers LDB1\href{https://paperpile.com/c/BKye4I/u71mp+n4P9o}{\textsuperscript{45,46}} (Car1 and Car2), Mt2 and Hmgb3\href{https://paperpile.com/c/BKye4I/PvEjT}{\textsuperscript{47}}. The logistic regression model finds both membrane-related genes (Car2, Minpp1\href{https://paperpile.com/c/BKye4I/84v9W}{\textsuperscript{48}}, Sphk1\href{https://paperpile.com/c/BKye4I/7Oc5L}{\textsuperscript{49}}, Tomm20, Tmem234) and the nuclear-localised genes Khsrp, Hmgb3\href{https://paperpile.com/c/BKye4I/71Jlg}{\textsuperscript{50}}, Mt1, Mt2, Nono\href{https://paperpile.com/c/BKye4I/SEtgs}{\textsuperscript{51}}, Sphk1\href{https://paperpile.com/c/BKye4I/rhD10}{\textsuperscript{52}}. During erythroid cell maturation, the cells need to prepare the membrane for metal binding and oxygen transportation, hence the expression of specific membrane proteins. 
\end{justify}\par

\vspace{\baselineskip}
\begin{justify}
At the early ETD stage, a different set of membrane genes are expressed (Rhd\href{https://paperpile.com/c/BKye4I/Jzx3W}{\textsuperscript{53}}, Dmtn\href{https://paperpile.com/c/BKye4I/mnK7O}{\textsuperscript{54}}, Lbr\href{https://paperpile.com/c/BKye4I/Xx5mf}{\textsuperscript{55}}, Slc4a1\href{https://paperpile.com/c/BKye4I/JDu3G}{\textsuperscript{56}}) to prepare the membrane for later metal binding. Cell cycle-related genes are still highly expressed at this stage, where Ccnd3, Mcm7, Birc5\href{https://paperpile.com/c/BKye4I/y5MHL}{\textsuperscript{57}} and Smc2 are highlighted by the logistic regression model. The haemoglobin gene Hbb-bt\href{https://paperpile.com/c/BKye4I/yQQHY}{\textsuperscript{58}} starts to become expressed at this stage. 
\end{justify}\par

\vspace{\baselineskip}
\begin{justify}
Finally, SCCAF predicts the erythrocyte development-related genes of Bpgm\href{https://paperpile.com/c/BKye4I/C2j9K}{\textsuperscript{59}}, Hba-a1, Hba-a2 and Hbb-bs as being upregulated at the late ETD stage, indicating the maturation of the red blood cells. Other metal-binding and oxygen transport-related genes such as Mkrn1\href{https://paperpile.com/c/BKye4I/0lrwd}{\textsuperscript{60}} and Rsad2\href{https://paperpile.com/c/BKye4I/Ix94B}{\textsuperscript{61}} are also expressed at this stage, while Alas2\href{https://paperpile.com/c/BKye4I/CJEmu}{\textsuperscript{62}} is a known erythrocyte marker that directly interacts with Hba-a1, Hba-a2, Hbb-bs and Hbb-bt. 
\end{justify}\par

\vspace{\baselineskip}
\setlength{\parskip}{0.0pt}
\begin{justify}
During the maturation process (\textbf{Figure 5h}), the cells in the CEP and early ETD stages are cycling and include more genes (larger cells), while the late ETD stage stops proliferating and the haemoglobins are expressed. In summary, these results demonstrate that the optimized clustering achieved by SCCAF shows biological relevance and can define key cell states within a continuous differentiation or development process. 
\end{justify}\par

\vspace{\baselineskip}
\subsection*{Discussion}
\addcontentsline{toc}{subsection}{Discussion}

\vspace{\baselineskip}
\begin{justify}
We have demonstrated that our proposed Single Cell Clustering Assessment Framework (SCCAF) can be used to automate the discovery of putative cell types from scRNA-seq data. Using 8 published datasets, we demonstrated that SCCAF restores the expert cell type assignments with high accuracy. As SCCAF associates each of the discovered cell types with a ranked list of feature genes defining this cell type, it effectively provides an initial cell type annotation. Moreover, the associated gene lists can then be used to derive a biologically meaningful cell-type annotation, either by a human expert or by applying one of the published automated reference-based methods\href{https://paperpile.com/c/BKye4I/z6JA+ia4B+mL7w+cWTh+YStx+1lv9+J2VQ+zaRi+kYlT}{\textsuperscript{22–30}}. We also demonstrate that in datasets providing temporal information about cell type/state progression, for instance, during cell differentiation, we can restore biologically meaningful cell states (\textbf{Figure 5}). 
\end{justify}\par

\vspace{\baselineskip}
\begin{justify}
We have compared the results from SCCAF with cell type annotations obtained from reference-based methods on previously published expert annotated datasets\href{https://paperpile.com/c/BKye4I/otnx+AXkh+eE5EA}{\textsuperscript{37–39}}. We show that SCCAF finds the correct cell groups, often outperforming the tested reference-based methods, even though it does not require a reference (\textbf{Figure S12-S14}). We have also shown that a model (\textit{i.e.} the weighted gene lists) trained by SCCAF on one dataset, can then be successfully applied to an independent dataset of the same tissue, to classify the cells automatically (\textbf{Figure 5}). Moreover, when a reference is only available for a different organism (for example comparing human and mouse brain), the model that SCCAF builds can be used for cross-species comparisons (\textbf{Figure S15}).
\end{justify}\par

\vspace{\baselineskip}
\begin{justify}
Cell type taxonomy is often presented as a hierarchy, but such representations are only approximations to biological reality, as they overlook the temporal aspect of cell development or progressive transitions, as well as cell state convergence\href{https://paperpile.com/c/BKye4I/2CQh+3Ubg}{\textsuperscript{63,64}}. Most scRNA-seq experiments typically study cells at a particular level of resolution, and SCCAF is designed to uncover the most appropriate $``$flat$"$  classifications for the particular level of resolution. Nevertheless, we show that if a dataset provides sufficient variances in gene expression, then by applying SCCAF iteratively, we can sometimes reconstruct parts of the underlying cell type hierarchy (\textbf{Figure S16}). (Fully automatic reconstruction of cell-type hierarchy is a challenge that is not addressed by SCCAF.) 
\end{justify}\par

\vspace{\baselineskip}
\begin{justify}
Importantly, the performance of our method is not significantly affected by the size of the clusters, and thus it can be used to discover rare cell types, as long as they are detected by the initial clustering. In the real-world datasets that we tested, the smallest original cluster was 7 cells, but more typically the smallest clusters are around 30 cells, a regime where SCCAAF works well (\textbf{suppl notes, suppl Figure S17}). There is always a compromise to be struck between the minimal number of cells in a group that is used to define a cell type \textit{versus} the potential introduction of noise. 
\end{justify}\par

\vspace{\baselineskip}
\begin{justify}
We have experimented with different types of classifiers in the machine-learning step of SCCAF and found that a regularised linear regression outperformed more sophisticated non-linear classifiers not only on simulated, but also on the real-world datasets (\textbf{Figure S2}). Although, it has been demonstrated previously that the linear regression classifiers can discriminate between cell types with good accuracy\href{https://paperpile.com/c/BKye4I/var3u}{\textsuperscript{31}}, this observation may appear surprising. One can argue that this is a consequence of how cell types are currently defined in the context of scRNA-seq data (\textit{i.e.}, via highly expressed marker genes), which is also consistent with how scRNA-seq data is usually clustered (either based on highly variable genes or PCs in expression space). It is an open question, whether there is an alternative biologically meaningful cell type definitions (for instance, based on decision trees), for which non-linear classifiers would outperform linear approaches. 
\end{justify}\par

\vspace{\baselineskip}
\begin{justify}
SCCAF has broad applications and can be implemented in a straightforward way. It is available as an open-source software package on GitHub (\href{https://github.com/SCCAF/sccaf}{\ul{https://github.com/SCCAF/sccaf}}) and as a Python package index (\href{https://pypi.org/project/SCCAF/}{\ul{https://pypi.org/project/SCCAF/}}), and is a Galaxy\href{https://paperpile.com/c/BKye4I/0WGt}{\textsuperscript{65}} tool in the Human Cell Atlas (https://humancellatlas.usegalaxy.eu). 
\end{justify}\par

\vspace{\baselineskip}

\vspace{\baselineskip}
\subsection*{Methods}
\addcontentsline{toc}{subsection}{Methods}
\subsubsection*{Machine learning-based self-projection}
\addcontentsline{toc}{subsubsection}{Machine learning-based self-projection}
\begin{justify}
As shown in \textbf{Figure 1a}, the expression profile and cluster assignment is first split into training and test set. If a cluster of cells includes more than 200 cells, 100 cells are randomly selected from a cluster and used as a training set, while all the rest of the cells are used as the test set. When a cluster contains between 10 and 200 cells, half of the cells are randomly selected as training set and the other half is used as the test set. Given that the algorithm uses machine learning and 5-fold cross-validation to build a classifier, if the cluster contains fewer than 10 cells, but at least 6 cells, our algorithm splits the cluster asymmetrically, so that 5 cells are included in the training set, while the remaining ones in the test set. Two parameters, the maximum number of cells used for training and the fraction of cells used for training, have been implemented to adjust the training/testing ratio. We train a multi-class machine learning classifier based on the training set. Taking the advantage of sklearn\href{https://paperpile.com/c/BKye4I/gZUmc}{\textsuperscript{66}}, we implement 5 different machine learning models: Logistic Regression, Random Forest, Gaussian Process Classifier, Support Vector Machine and Decision Tree. In the case of logistic regression, we use ‘L1’ regularization to avoid overfitting. 5-fold cross-validation, which is implemented in sklearn.model\_selection, is applied to the training set. The average value of the cross-validation accuracies was used as the accuracy of the cross-validation. The model trained on the training set is then applied to the test set and the predicted results of the test set are compared with its original clustering. \textit{Self-projection accuracy} is defined as the percentage of correctly predicted cells in the test set. It can be considered as a metric to assess the reliability of the clustering. 
\end{justify}\par

\vspace{\baselineskip}
\begin{justify}
The Precision-Recall curves were calculated by the sklearn.metrics.precision\_recall\_curve function. The Receiver Operating Characteristic (ROC) curves were calculated on the relationship between False Positive Rate (FPR) and True Positive Rate (TPR) using the sklearn.metrics.roc\_curve function. Area Under Curve (AUC), calculated by the sklearn.metrics.auc function, is used as a metric to benchmark the accuracy of a prediction model. 
\end{justify}\par

\vspace{\baselineskip}
\subsubsection*{Confusion matrix-directed cluster optimization}
\addcontentsline{toc}{subsubsection}{Confusion matrix-directed cluster optimization}
\begin{justify}
\textbf{Figure S18 }describes\textbf{ }the whole workflow of SCCAF optimization. SCCAF uses the \textit{confusion matrix\href{https://paperpile.com/c/BKye4I/7oPv}{\textsuperscript{67}}} obtained from the test set to identify cell clusters that are likely to include cells of the same type. The confusion matrix \textit{C} is an \textit{n} x \textit{n} matrix, where \textit{n} is the number of clusters, and the elements of the matrix \textit{c\textsubscript{i,j}} represent the number of cells in cluster \textit{j }but\textit{ predicted} as belonging to the cluster \textit{i. Thus} \textit{c\textsubscript{i,i}} is the number of cells in the cluster \textit{i} also predicted as belonging to \textit{i}. We refer to \textit{c\textsubscript{i,j}} as the \textit{confusion rate}. Confusion matrix is calculated by the sklearn.metrics.confusion\_matrix function based on the clustering assignment of the test set and the predicted clustering from the machine learning model. 
\end{justify}\par

\vspace{\baselineskip}
\begin{justify}
This confusion matrix is then normalized. We define the \textit{normalized confusion rate} \textit{r(i,j)} between clusters \textit{i} and \textit{j}, as the maximum of ratios of misclassified and correctly classified cells as follows: 
\end{justify}\par

\vspace{\baselineskip}
\begin{justify}
 \( r \left( i,j \right) =max \{ \frac{C_{i,j}}{C_{i,i}},\frac{C_{j,i}}{C_{j,j}} \}  \) 
\end{justify}\par

\vspace{\baselineskip}

\vspace{\baselineskip}
\begin{justify}
Intuitively, \textit{r(i,j)} accounts for the confusion rate relative to the correctly assigned cell numbers in a cluster. For example, an \textit{r}=0.3 means 30$\%$  of the cells are confused between clusters. High normalised confusion rate indicates that the clusters \textit{i} and \textit{j} are likely to represent the cells of the same type. 
\end{justify}\par

\vspace{\baselineskip}
\begin{justify}
The normalized confusion matrix is then binarized into a \textit{connection matrix} by a threshold of normalized confusion rate. The threshold is defined as: \textit{r\textsubscript{threshold}} = \textit{max}$ \{ $ \textit{r(i,j)}$ \} $  - 0.01, which is 1$\%$  lower than the maximum normalized confusion rate of the current clustering. An example of the normalization which corresponds to the simulated data in Figure 2 is shown in \textbf{Figure S19}. The connection matrix is then converted into a connection graph using the python igraph library implemented in SCANPY. Merged groups are obtained by applying the Louvain clustering algorithm on the generated connection graph. The whole cluster merging optimization process is iteratively performed until the pre-set self-projection accuracy is achieved. 
\end{justify}\par

\vspace{\baselineskip}
\begin{justify}
During the clustering optimization, graphical plots can be output to show the clustering assignments, the self-projection result, the Precision-Recall curves, the confusion matrix and normalized confusion matrices. All graphical plots were generated by SCANPY and matplotlib\href{https://paperpile.com/c/BKye4I/gfgWA}{\textsuperscript{68}}. 
\end{justify}\par

\subsubsection*{Data simulation}
\addcontentsline{toc}{subsubsection}{Data simulation}
\begin{justify}
In the theory test, two types of data simulations were used: a multivariate normal simulation and the Splatter\href{https://paperpile.com/c/BKye4I/fFIer}{\textsuperscript{32}} simulation. 
\end{justify}\par

\paragraph*{Multivariate normal simulation}
\addcontentsline{toc}{paragraph}{Multivariate normal simulation}
\begin{justify}
We first simulated data with a multivariate normal distribution using the $``$multivariate\_normal$"$  function in Scipy\href{https://paperpile.com/c/BKye4I/DWXyj}{\textsuperscript{69}} in the same approach as\href{https://paperpile.com/c/BKye4I/ZLP4X}{\textsuperscript{70}}. Gene expression profiles x = (x\textsubscript{1}, ..., x\textsubscript{N}) for each gene follows a normal distribution:
\end{justify}\par


\begin{figure}[H]
	\begin{Center}
		X ~ N(myu, Sigma)
	\end{Center}
\end{figure}


\par

\begin{justify}
The background genes have an average expression value 1 and the marker genes have mean expression value 3. Each cell state includes a random number of marker genes between 10-20, while 100 background genes were added besides all the marker genes. 
\end{justify}\par

\paragraph*{Splatter simulation}
\addcontentsline{toc}{paragraph}{Splatter simulation}
\begin{justify}
We use the splatter\href{https://paperpile.com/c/BKye4I/fFIer}{\textsuperscript{32}} program to simulate data in a more realistic way. Default parameters for simulation were estimated using the \textit{splatEstimate} function. The differential expression parameters were set to: group.prob=0.5, de.prob=1, de.facLoc=0.1, de.facScale = 0.5, nGenes = 200. The function \textit{splatSimulateGroups} was used to simulate the data. 
\end{justify}\par

\subsubsection*{Extracting marker genes based on the logistic regression model}
\addcontentsline{toc}{subsubsection}{Extracting marker genes based on the logistic regression model}
\begin{justify}
We get the weight for each gene in each of the cell clusters from the $``$.coef\_$"$  parameter of the logistic regression model. Only positive weights were extracted. The weights are then sorted in decreasing order, while the top 20 ranked genes are extracted as potential feature genes. 
\end{justify}\par

\vspace{\baselineskip}
\subsubsection*{Datasets}
\addcontentsline{toc}{subsubsection}{Datasets}
\subsubsection*{Data sets and processing}
\addcontentsline{toc}{subsubsection}{Data sets and processing}
\paragraph*{Shekhar (Mouse Retina) data}
\addcontentsline{toc}{paragraph}{Shekhar (Mouse Retina) data}
\begin{justify}
We downloaded the digital gene expression data from GEO accession \href{ftp://ftp.ncbi.nlm.nih.gov/geo/series/GSE81nnn/GSE81904/suppl/GSE81904_bipolar_data_Cell2016.Rdata.gz}{GSE65785} which is referenced in ref \href{https://paperpile.com/c/BKye4I/d4ibg}{\textsuperscript{13}}. Cells with more than 10$\%$  mitochondrial contents were excluded. Cells from Bipolar5 and Bipolar6 are assigned as batch 2, while all other cells are assigned as batch 1. The COMBAT function from svaseq\href{https://paperpile.com/c/BKye4I/qMDKz}{\textsuperscript{73}} was used to correct the batch effect. Cells annotated as Doublets/Contaminants are excluded from the analysis. 100 principal components were used to analyze the cell clusters. 
\end{justify}\par

\paragraph*{Zeisel data}
\addcontentsline{toc}{paragraph}{Zeisel data}
\begin{justify}
From the Hemberg lab scRNA-seq datasets website (\href{https://hemberg-lab.github.io/scRNA.seq.datasets/}{\ul{https://hemberg-lab.github.io/scRNA.seq.datasets/}}), we downloaded the count matrix together with its annotation of mouse cortex data from \href{https://paperpile.com/c/BKye4I/Gfzwj}{\textsuperscript{35}}. We filtered cells expressing fewer than 200 genes and genes expressed in fewer than 3 cells. We use the top 2000 variable genes to represent the variance of the dataset based on the standard deviation of the genes, similar to the approach in the BISCUIT publication \href{https://paperpile.com/c/BKye4I/Bs9JL}{\textsuperscript{74}}. 
\end{justify}\par

\paragraph*{Segerstolpe data}
\addcontentsline{toc}{paragraph}{Segerstolpe data}
\begin{justify}
We downloaded the processed expression matrix and cell-type annotation from ArrayExpress\href{https://paperpile.com/c/BKye4I/nZodk}{\textsuperscript{75}} accession \href{https://www.ebi.ac.uk/arrayexpress/files/E-MTAB-5061/E-MTAB-5061.processed.1.zip}{\ul{E-MTAB-5061}} which corresponds to the data in \href{https://paperpile.com/c/BKye4I/dzVJ6}{\textsuperscript{15}}. We remove the uncertain cells annotated as ‘not applicable’, ‘unclassified endocrine cell’, ‘unclassified cell’,‘co-expression cell’, ‘MHC class II cell’ as well as the cell types of fewer than 10 cells (mast cell and epsilon cell). Highly variable genes were selected based on the mean expression and dispersions. 
\end{justify}\par

\paragraph*{Baron data}
\addcontentsline{toc}{paragraph}{Baron data}
\begin{justify}
The count matrices together with their annotations for mouse and human were downloaded from GEO accession GSE84133 as ref \href{https://paperpile.com/c/BKye4I/vAx8W}{\textsuperscript{14}}. Cells with fewer than 200 genes or more than 12000 cells were removed from consideration, while genes expressed in fewer than 3 cells were removed. The linear regression function from NaiveDE (\href{https://github.com/Teichlab/NaiveDE}{\ul{https://github.com/Teichlab/NaiveDE}}) was used to regress out donor effect as well as the technical variances from the number of genes and number of counts. 
\end{justify}\par

\paragraph*{Hrvatin data}
\addcontentsline{toc}{paragraph}{Hrvatin data}
\begin{justify}
We downloaded the raw count matrix and the cell type annotation from GEO accession GSE102827 as ref \href{https://paperpile.com/c/BKye4I/eE5EA}{\textsuperscript{37}}. We filter out cells of fewer than 200 genes and genes expressed in fewer than 3 cells. Cells annotated as ‘nan’ were removed from analysis. Highly variable genes were selected based on mean expression and dispersion. 
\end{justify}\par

\paragraph*{Tusi data}
\addcontentsline{toc}{paragraph}{Tusi data}
\begin{justify}
The count matrix for ref \href{https://paperpile.com/c/BKye4I/RdCh9}{\textsuperscript{41}} was downloaded from GEO accession \href{https://www.ncbi.nlm.nih.gov/geo/query/acc.cgi?acc=GSE89754}{GSE89754}. We remove the cells from ‘basal\_bm1’ to avoid dealing with batch effects. 
\end{justify}\par

\paragraph*{Giladi data}
\addcontentsline{toc}{paragraph}{Giladi data}
\begin{justify}
The count matrix for ref \href{https://paperpile.com/c/BKye4I/DmgXV}{\textsuperscript{42}} was downloaded from GEO accession GSE92575. Only cells without any treatment were used in this analysis. ERCC spikes-ins were removed from the analysis, while batch effect related to the $``$Seq\_batch\_ID$"$  was regressed out by COMBAT. 
\end{justify}\par

\paragraph*{Human Brain data}
\addcontentsline{toc}{paragraph}{Human Brain data}
\begin{justify}
Human brain single nucleic-Seq datasets for Middle Temporal Gyrus (MTG, 15,928 nuclei), Primary Visual Cortex (VISp, 8,998 nuclei), Anterior Cingulate Cortex (ACC, 7,283 nuclei) and Lateral Geniculate (LGN, 1,576 nuclei) were downloaded from Allen Cell Types Database\href{https://paperpile.com/c/BKye4I/oyhFn}{\textsuperscript{76}}. We filter out cells of fewer than 200 genes and genes expressed in fewer than 3 cells. 
\end{justify}\par

\setlength{\parskip}{12.0pt}
\paragraph*{MacParland data}
\addcontentsline{toc}{paragraph}{MacParland data}
\begin{justify}
Human liver data from MacParland et al.\href{https://paperpile.com/c/BKye4I/ea2W}{\textsuperscript{72}} was used as a reference dataset for the reference-based cell-type annotation methods. Data was downloaded from GEO accession \href{https://www.ncbi.nlm.nih.gov/geo/query/acc.cgi?acc=GSE115469}{GSE115469}. The expression matrix was log-transformed without pre-processing. 
\end{justify}\par

\paragraph*{Aizarani data}
\addcontentsline{toc}{paragraph}{Aizarani data}
\begin{justify}
Human liver data from Aizarani et al. \href{https://paperpile.com/c/BKye4I/otnx}{\textsuperscript{38}}, including the count matrix and the cluster assignment, was downloaded from the GEO accession \href{https://www.ncbi.nlm.nih.gov/geo/query/acc.cgi?acc=GSE124395}{GSE124395} and cell types were annotated according to the clusters in Figure 1 of the reference paper\href{https://paperpile.com/c/BKye4I/otnx}{\textsuperscript{38}}. Cells expression fewer than 200 genes and genes expressed in fewer than 20 cells were excluded. Batch information was inferred from the cell names and the batch effect was regressed out using the ‘regress\_out’ function in SCANPY. 
\end{justify}\par

\paragraph*{Tasic datasets}
\addcontentsline{toc}{paragraph}{Tasic datasets}
\begin{justify}
The mouse cortex datasets from Tasic et al. \href{https://paperpile.com/c/BKye4I/QNup+AXkh}{\textsuperscript{39,71}} were used to benchmark the reference-based cell-type annotation methods. Tasic2016 data \href{https://paperpile.com/c/BKye4I/QNup}{\textsuperscript{71}} was used as a reference, while Tasic2018 data \href{https://paperpile.com/c/BKye4I/AXkh}{\textsuperscript{39}} was used as a benchmark. The count matrix and cell-type annotation of the Tasic2016 data were downloaded from GEO accession \href{https://www.ncbi.nlm.nih.gov/geo/query/acc.cgi?acc=GSE71585}{GSE71585}. We exclude Cells expression fewer than 200 genes and genes expressed in fewer than 20 cells. The exon counts matrix and metadata table of Tasic2018 data were downloaded from GEO accession \href{https://www.ncbi.nlm.nih.gov/geo/query/acc.cgi?acc=GSE115746}{GSE115746}. Cells expression fewer than 200 genes and genes expressed in fewer than 3 cells were excluded.
\end{justify}\par

\begin{justify}
All expression data and metadata were imported into the SCANPY\href{https://paperpile.com/c/BKye4I/VJJ9r}{\textsuperscript{8}} python class and saved as HDF5 files. All the data pre-processing steps are saved in Jupiter notebooks and are available at GitHub (\href{https://github.com/SCCAF/sccaf_example}{\ul{https://github.com/SCCAF/sccaf\_example}}). 
\end{justify}\par

\vspace{\baselineskip}
\subsubsection*{The SCANPY analysis Workflow}
\addcontentsline{toc}{subsubsection}{The SCANPY analysis Workflow}
\begin{justify}
All the datasets were visualized using a common analysis process based on a standard SCANPY workflow, which includes 7 steps of processing: 1) normalize the data to 10,000 counts per cell; 2) identify highly variable genes based mean expressions of the genes and normalized dispersions of the genes; 3) log transform the data and scale to unit variance and zero mean; 4) dimension reduction by PCA; 5) measure the nearest neighbour graph based on the top 15 PCs; 6) dimension reduction by tSNE and UMAP and 7) Louvain clustering based on the nearest neighbour graph. 
\end{justify}\par

\subsubsection*{Reference-based cell-type annotation methods}
\addcontentsline{toc}{subsubsection}{Reference-based cell-type annotation methods}
\begin{justify}
To benchmark SCCAF, we compare with the reference-based cell-type annotation methods, including logistic regression, SingleR\href{https://paperpile.com/c/BKye4I/YStx}{\textsuperscript{26}}, singleCellNet \href{https://paperpile.com/c/BKye4I/1lv9}{\textsuperscript{27}}, moana\href{https://paperpile.com/c/BKye4I/J2VQ}{\textsuperscript{28}}, ACTINN\href{https://paperpile.com/c/BKye4I/zaRi}{\textsuperscript{29}}, scClassify\href{https://paperpile.com/c/BKye4I/kYlT}{\textsuperscript{30}} and CHETAH\href{https://paperpile.com/c/BKye4I/z6JA}{\textsuperscript{22}}. All the methods read the data from the SCANPY hdf5 files generated using the preprocessing notebooks. All the scripts used to run these programs are available at GitHub (\href{https://github.com/SCCAF/sccaf_example}{\ul{https://github.com/SCCAF/sccaf\_example}}).
\end{justify}\par

\vspace{\baselineskip}
\subsubsection*{Data Availability}
\addcontentsline{toc}{subsubsection}{Data Availability}
The datasets together with the accession codes are listed below:\par

\vspace{\baselineskip}


\begin{table}[H]
 			\centering
\begin{tabular}{p{1.37in}p{1.37in}p{1.37in}p{1.37in}}
\hline
\multicolumn{1}{|p{1.37in}}{Name} & 
\multicolumn{1}{|p{1.37in}}{Tissue} & 
\multicolumn{1}{|p{1.37in}}{Accession} & 
\multicolumn{1}{|p{1.37in}|}{Reference} \\
\hhline{----}
\multicolumn{1}{|p{1.37in}}{Baron et al.} & 
\multicolumn{1}{|p{1.37in}}{Pancreas} & 
\multicolumn{1}{|p{1.37in}}{GSE84133} & 
\multicolumn{1}{|p{1.37in}|}{\href{https://paperpile.com/c/BKye4I/vAx8W}{\textsuperscript{14}}} \\
\hhline{----}
\multicolumn{1}{|p{1.37in}}{Zeisel et al.} & 
\multicolumn{1}{|p{1.37in}}{Cortex} & 
\multicolumn{1}{|p{1.37in}}{GSE60361} & 
\multicolumn{1}{|p{1.37in}|}{\href{https://paperpile.com/c/BKye4I/Gfzwj}{\textsuperscript{35}}} \\
\hhline{----}
\multicolumn{1}{|p{1.37in}}{Shekhar et al.} & 
\multicolumn{1}{|p{1.37in}}{Retinal Bipolar Neurons} & 
\multicolumn{1}{|p{1.37in}}{GSE81904} & 
\multicolumn{1}{|p{1.37in}|}{\href{https://paperpile.com/c/BKye4I/d4ibg}{\textsuperscript{13}}} \\
\hhline{----}
\multicolumn{1}{|p{1.37in}}{Segerstolpe et al.} & 
\multicolumn{1}{|p{1.37in}}{Pancreatic Islets} & 
\multicolumn{1}{|p{1.37in}}{E-MTAB-5061} & 
\multicolumn{1}{|p{1.37in}|}{\href{https://paperpile.com/c/BKye4I/dzVJ6}{\textsuperscript{15}}} \\
\hhline{----}
\multicolumn{1}{|p{1.37in}}{Hrvatin et al. } & 
\multicolumn{1}{|p{1.37in}}{Visual cortex} & 
\multicolumn{1}{|p{1.37in}}{GSE102827} & 
\multicolumn{1}{|p{1.37in}|}{\href{https://paperpile.com/c/BKye4I/eE5EA}{\textsuperscript{37}}} \\
\hhline{----}
\multicolumn{1}{|p{1.37in}}{Tusi et al.} & 
\multicolumn{1}{|p{1.37in}}{haematopoiesis} & 
\multicolumn{1}{|p{1.37in}}{GSE89754} & 
\multicolumn{1}{|p{1.37in}|}{\href{https://paperpile.com/c/BKye4I/RdCh9}{\textsuperscript{41}}} \\
\hhline{----}
\multicolumn{1}{|p{1.37in}}{Giladi et al.} & 
\multicolumn{1}{|p{1.37in}}{haematopoiesis} & 
\multicolumn{1}{|p{1.37in}}{GSE92575} & 
\multicolumn{1}{|p{1.37in}|}{\href{https://paperpile.com/c/BKye4I/DmgXV}{\textsuperscript{42}}} \\
\hhline{----}
\multicolumn{1}{|p{1.37in}}{Tasic et al. 2016} & 
\multicolumn{1}{|p{1.37in}}{Cortex} & 
\multicolumn{1}{|p{1.37in}}{\href{https://www.ncbi.nlm.nih.gov/geo/query/acc.cgi?acc=GSE71585}{GSE71585}} & 
\multicolumn{1}{|p{1.37in}|}{\href{https://paperpile.com/c/BKye4I/QNup}{\textsuperscript{71}}} \\
\hhline{----}
\multicolumn{1}{|p{1.37in}}{Tasic et al. 2018} & 
\multicolumn{1}{|p{1.37in}}{Cortex} & 
\multicolumn{1}{|p{1.37in}}{\href{https://www.ncbi.nlm.nih.gov/geo/query/acc.cgi?acc=GSE115746}{GSE115746}} & 
\multicolumn{1}{|p{1.37in}|}{\href{https://paperpile.com/c/BKye4I/AXkh}{\textsuperscript{39}}} \\
\hhline{----}
\multicolumn{1}{|p{1.37in}}{Aizarani et al.} & 
\multicolumn{1}{|p{1.37in}}{Liver} & 
\multicolumn{1}{|p{1.37in}}{\href{https://www.ncbi.nlm.nih.gov/geo/query/acc.cgi?acc=GSE124395}{GSE124395}} & 
\multicolumn{1}{|p{1.37in}|}{\href{https://paperpile.com/c/BKye4I/otnx}{\textsuperscript{38}}} \\
\hhline{----}
\multicolumn{1}{|p{1.37in}}{MacParland et al. } & 
\multicolumn{1}{|p{1.37in}}{Liver} & 
\multicolumn{1}{|p{1.37in}}{\href{https://www.ncbi.nlm.nih.gov/geo/query/acc.cgi?acc=GSE115469}{GSE115469}} & 
\multicolumn{1}{|p{1.37in}|}{\href{https://paperpile.com/c/BKye4I/ea2W}{\textsuperscript{72}}} \\
\hhline{----}

\end{tabular}
 \end{table}


\vspace{\baselineskip}
\subsection*{Acknowledgements}
\addcontentsline{toc}{subsection}{Acknowledgements}
\setlength{\parskip}{0.0pt}
\begin{justify}
The authors would like to thank all members of the Teichmann lab and Brazma lab for helpful discussions. Z.M. is supported by a Single Cell Gene Expression Atlas grant from the Wellcome Trust (nr. 108437/Z/15/Z). 
\end{justify}\par

\tab \tab \tab \tab \tab 
\vspace{\baselineskip}\begin{justify}
Author contributions: ZM conceived the method, implemented the algorithm and website, conducted the analyses, created the figures and contributed to the manuscript. PM, NH and IP packed the algorithm and implemented as a Galaxy tool. AB and SAT supervised the work and contributed to the manuscript. 
\end{justify}\par

\tab \tab \tab \tab \tab 
\vspace{\baselineskip}\begin{justify}
Competing interests: none declared.
\end{justify}\par

\vspace{\baselineskip}
\textbf{Code availability}\par

\begin{justify}
An open-source implementation of SCCAF is available at GitHub (\href{https://github.com/SCCAF/sccaf}{\ul{https://github.com/SCCAF/sccaf}}) and (\href{https://github.com/functional-Genomics/SCCAF}{\ul{https://github.com/functional-Genomics/SCCAF}}). The release includes tutorials and example vignettes for reproducing the presented analyses, as well as all preprocessed data sets considered in this study. The software version used to generate the results presented in this paper is also available as \textbf{Supplementary Software}. SCCAF is also accessible from Python package index (\href{https://pypi.org/project/SCCAF/}{\ul{https://pypi.org/project/SCCAF/}}). And it is implemented as a Galaxy tool in the Human Cell Atlas (\href{https://humancellatlas.usegalaxy.eu/}{\ul{https://humancellatlas.usegalaxy.eu/}}). The SCCAF Galaxy modules are available to install with a few clicks on any Galaxy instance through the main Galaxy Toolshed at \href{https://toolshed.g2.bx.psu.edu/view/ebi-gxa/suite_sccaf/}{\ul{https://toolshed.g2.bx.psu.edu/view/ebi-gxa/suite\_sccaf/}}.
\end{justify}\par

\subsection*{References}
\addcontentsline{toc}{subsection}{References}

\vspace{\baselineskip}
\setstretch{2.0}
\begin{adjustwidth}{0.31in}{0.0in}
\begin{FlushLeft}
{\fontsize{11pt}{13.2pt}\selectfont 1.\tab \href{http://paperpile.com/b/BKye4I/YyFyV}{Hooke, R. $\&$  Jo Martyn And. Micrographia, or, Some physiological descriptions of minute bodies made by magnifying glasses :with observations and inquiries thereupon /by R. Hooke . (1665). doi:}\href{http://dx.doi.org/10.5962/bhl.title.904}{10.5962/bhl.title.904}\par}
\end{FlushLeft}\par

\end{adjustwidth}

\begin{adjustwidth}{0.31in}{0.0in}
\begin{FlushLeft}
{\fontsize{11pt}{13.2pt}\selectfont 2.\tab \href{http://paperpile.com/b/BKye4I/qmBan}{Arendt, D. }\href{http://paperpile.com/b/BKye4I/qmBan}{\textit{et al.}\href{http://paperpile.com/b/BKye4I/qmBan}{} The origin and evolution of cell types. }\href{http://paperpile.com/b/BKye4I/qmBan}{\textit{Nat. Rev. Genet.}\href{http://paperpile.com/b/BKye4I/qmBan}{} }\href{http://paperpile.com/b/BKye4I/qmBan}{\textbf{17}\href{http://paperpile.com/b/BKye4I/qmBan}{}, 744 (2016).}\par}
\end{FlushLeft}\par

\end{adjustwidth}

\begin{adjustwidth}{0.31in}{0.0in}
\begin{FlushLeft}
{\fontsize{11pt}{13.2pt}\selectfont 3.\tab \href{http://paperpile.com/b/BKye4I/I8LQd}{Nagasawa, T. Microenvironmental niches in the bone marrow required for B-cell development. }\href{http://paperpile.com/b/BKye4I/I8LQd}{\textit{Nat. Rev. Immunol.}\href{http://paperpile.com/b/BKye4I/I8LQd}{} }\href{http://paperpile.com/b/BKye4I/I8LQd}{\textbf{6}\href{http://paperpile.com/b/BKye4I/I8LQd}{}, 107–116 (2006).}\par}
\end{FlushLeft}\par

\end{adjustwidth}

\begin{adjustwidth}{0.31in}{0.0in}
\begin{FlushLeft}
{\fontsize{11pt}{13.2pt}\selectfont 4.\tab \href{http://paperpile.com/b/BKye4I/6OnXU}{Pouyan, M. B., Jindal, V., Birjandtalab, J. $\&$  Nourani, M. Single and multi-subject clustering of flow cytometry data for cell-type identification and anomaly detection. }\href{http://paperpile.com/b/BKye4I/6OnXU}{\textit{BMC Med. Genomics}\href{http://paperpile.com/b/BKye4I/6OnXU}{} }\href{http://paperpile.com/b/BKye4I/6OnXU}{\textbf{9 Suppl 2}\href{http://paperpile.com/b/BKye4I/6OnXU}{}, 41 (2016).}\par}
\end{FlushLeft}\par

\end{adjustwidth}

\begin{adjustwidth}{0.31in}{0.0in}
\begin{FlushLeft}
{\fontsize{11pt}{13.2pt}\selectfont 5.\tab \href{http://paperpile.com/b/BKye4I/8amOZ}{Trapnell, C. Defining cell types and states with single-cell genomics. }\href{http://paperpile.com/b/BKye4I/8amOZ}{\textit{Genome Res.}\href{http://paperpile.com/b/BKye4I/8amOZ}{} }\href{http://paperpile.com/b/BKye4I/8amOZ}{\textbf{25}\href{http://paperpile.com/b/BKye4I/8amOZ}{}, 1491–1498 (2015).}\par}
\end{FlushLeft}\par

\end{adjustwidth}

\begin{adjustwidth}{0.31in}{0.0in}
\begin{FlushLeft}
{\fontsize{11pt}{13.2pt}\selectfont 6.\tab \href{http://paperpile.com/b/BKye4I/g1tAJ}{Rozenblatt-Rosen, O., Stubbington, M. J. T., Regev, A. $\&$  Teichmann, S. A. The Human Cell Atlas: from vision to reality. }\href{http://paperpile.com/b/BKye4I/g1tAJ}{\textit{Nature}\href{http://paperpile.com/b/BKye4I/g1tAJ}{} }\href{http://paperpile.com/b/BKye4I/g1tAJ}{\textbf{550}\href{http://paperpile.com/b/BKye4I/g1tAJ}{}, 451–453 (2017).}\par}
\end{FlushLeft}\par

\end{adjustwidth}

\begin{adjustwidth}{0.31in}{0.0in}
\begin{FlushLeft}
{\fontsize{11pt}{13.2pt}\selectfont 7.\tab \href{http://paperpile.com/b/BKye4I/rvY0t}{Regev, A. }\href{http://paperpile.com/b/BKye4I/rvY0t}{\textit{et al.}\href{http://paperpile.com/b/BKye4I/rvY0t}{} The Human Cell Atlas. }\href{http://paperpile.com/b/BKye4I/rvY0t}{\textit{Elife}\href{http://paperpile.com/b/BKye4I/rvY0t}{} }\href{http://paperpile.com/b/BKye4I/rvY0t}{\textbf{6}\href{http://paperpile.com/b/BKye4I/rvY0t}{}, (2017).}\par}
\end{FlushLeft}\par

\end{adjustwidth}

\begin{adjustwidth}{0.31in}{0.0in}
\begin{FlushLeft}
{\fontsize{11pt}{13.2pt}\selectfont 8.\tab \href{http://paperpile.com/b/BKye4I/VJJ9r}{Wolf, F. A., Angerer, P. $\&$  Theis, F. J. SCANPY: large-scale single-cell gene expression data analysis. }\href{http://paperpile.com/b/BKye4I/VJJ9r}{\textit{Genome Biol.}\href{http://paperpile.com/b/BKye4I/VJJ9r}{} }\href{http://paperpile.com/b/BKye4I/VJJ9r}{\textbf{19}\href{http://paperpile.com/b/BKye4I/VJJ9r}{}, 15 (2018).}\par}
\end{FlushLeft}\par

\end{adjustwidth}

\begin{adjustwidth}{0.31in}{0.0in}
\begin{FlushLeft}
{\fontsize{11pt}{13.2pt}\selectfont 9.\tab \href{http://paperpile.com/b/BKye4I/cthgZ}{Butler, A., Hoffman, P., Smibert, P., Papalexi, E. $\&$  Satija, R. Integrating single-cell transcriptomic data across different conditions, technologies, and species. }\href{http://paperpile.com/b/BKye4I/cthgZ}{\textit{Nat. Biotechnol.}\href{http://paperpile.com/b/BKye4I/cthgZ}{} }\href{http://paperpile.com/b/BKye4I/cthgZ}{\textbf{36}\href{http://paperpile.com/b/BKye4I/cthgZ}{}, 411–420 (2018).}\par}
\end{FlushLeft}\par

\end{adjustwidth}

\begin{adjustwidth}{0.31in}{0.0in}
\begin{FlushLeft}
{\fontsize{11pt}{13.2pt}\selectfont 10.\tab \href{http://paperpile.com/b/BKye4I/BggCD}{Traag, V., Waltman, L. $\&$  van Eck, N. J. From Louvain to Leiden: guaranteeing well-connected communities. (2018).}\par}
\end{FlushLeft}\par

\end{adjustwidth}

\begin{adjustwidth}{0.31in}{0.0in}
\begin{FlushLeft}
{\fontsize{11pt}{13.2pt}\selectfont 11.\tab \href{http://paperpile.com/b/BKye4I/lCRQo}{Lloyd, S. Least squares quantization in PCM. }\href{http://paperpile.com/b/BKye4I/lCRQo}{\textit{IEEE Trans. Inf. Theory}\href{http://paperpile.com/b/BKye4I/lCRQo}{} }\href{http://paperpile.com/b/BKye4I/lCRQo}{\textbf{28}\href{http://paperpile.com/b/BKye4I/lCRQo}{}, 129–137 (1982).}\par}
\end{FlushLeft}\par

\end{adjustwidth}

\begin{adjustwidth}{0.31in}{0.0in}
\begin{FlushLeft}
{\fontsize{11pt}{13.2pt}\selectfont 12.\tab \href{http://paperpile.com/b/BKye4I/Jazbe}{Blondel, V. D., Guillaume, J.-L., Lambiotte, R. $\&$  Lefebvre, E. Fast unfolding of communities in large networks. }\href{http://paperpile.com/b/BKye4I/Jazbe}{\textit{J. Stat. Mech: Theory Exp.}\href{http://paperpile.com/b/BKye4I/Jazbe}{} }\href{http://paperpile.com/b/BKye4I/Jazbe}{\textbf{2008}\href{http://paperpile.com/b/BKye4I/Jazbe}{}, P10008 (2008).}\par}
\end{FlushLeft}\par

\end{adjustwidth}

\begin{adjustwidth}{0.31in}{0.0in}
\begin{FlushLeft}
{\fontsize{11pt}{13.2pt}\selectfont 13.\tab \href{http://paperpile.com/b/BKye4I/d4ibg}{Shekhar, K. }\href{http://paperpile.com/b/BKye4I/d4ibg}{\textit{et al.}\href{http://paperpile.com/b/BKye4I/d4ibg}{} Comprehensive Classification of Retinal Bipolar Neurons by Single-Cell Transcriptomics. }\href{http://paperpile.com/b/BKye4I/d4ibg}{\textit{Cell}\href{http://paperpile.com/b/BKye4I/d4ibg}{} }\href{http://paperpile.com/b/BKye4I/d4ibg}{\textbf{166}\href{http://paperpile.com/b/BKye4I/d4ibg}{}, 1308–1323.e30 (2016).}\par}
\end{FlushLeft}\par

\end{adjustwidth}

\begin{adjustwidth}{0.31in}{0.0in}
\begin{FlushLeft}
{\fontsize{11pt}{13.2pt}\selectfont 14.\tab \href{http://paperpile.com/b/BKye4I/vAx8W}{Baron, M. }\href{http://paperpile.com/b/BKye4I/vAx8W}{\textit{et al.}\href{http://paperpile.com/b/BKye4I/vAx8W}{} A Single-Cell Transcriptomic Map of the Human and Mouse Pancreas Reveals Inter- and Intra-cell Population Structure. }\href{http://paperpile.com/b/BKye4I/vAx8W}{\textit{Cell Systems}\href{http://paperpile.com/b/BKye4I/vAx8W}{} }\href{http://paperpile.com/b/BKye4I/vAx8W}{\textbf{3}\href{http://paperpile.com/b/BKye4I/vAx8W}{}, 346–360.e4 (2016).}\par}
\end{FlushLeft}\par

\end{adjustwidth}

\begin{adjustwidth}{0.31in}{0.0in}
\begin{FlushLeft}
{\fontsize{11pt}{13.2pt}\selectfont 15.\tab \href{http://paperpile.com/b/BKye4I/dzVJ6}{Segerstolpe, Å. }\href{http://paperpile.com/b/BKye4I/dzVJ6}{\textit{et al.}\href{http://paperpile.com/b/BKye4I/dzVJ6}{} Single-Cell Transcriptome Profiling of Human Pancreatic Islets in Health and Type 2 Diabetes. }\href{http://paperpile.com/b/BKye4I/dzVJ6}{\textit{Cell Metab.}\href{http://paperpile.com/b/BKye4I/dzVJ6}{} }\href{http://paperpile.com/b/BKye4I/dzVJ6}{\textbf{24}\href{http://paperpile.com/b/BKye4I/dzVJ6}{}, 593–607 (2016).}\par}
\end{FlushLeft}\par

\end{adjustwidth}

\begin{adjustwidth}{0.31in}{0.0in}
\begin{FlushLeft}
{\fontsize{11pt}{13.2pt}\selectfont 16.\tab \href{http://paperpile.com/b/BKye4I/Ve8gd}{Han, X. }\href{http://paperpile.com/b/BKye4I/Ve8gd}{\textit{et al.}\href{http://paperpile.com/b/BKye4I/Ve8gd}{} Mapping the Mouse Cell Atlas by Microwell-Seq. }\href{http://paperpile.com/b/BKye4I/Ve8gd}{\textit{Cell}\href{http://paperpile.com/b/BKye4I/Ve8gd}{} }\href{http://paperpile.com/b/BKye4I/Ve8gd}{\textbf{173}\href{http://paperpile.com/b/BKye4I/Ve8gd}{}, 1307 (2018).}\par}
\end{FlushLeft}\par

\end{adjustwidth}

\begin{adjustwidth}{0.31in}{0.0in}
\begin{FlushLeft}
{\fontsize{11pt}{13.2pt}\selectfont 17.\tab \href{http://paperpile.com/b/BKye4I/ovLNj}{Tabula Muris Consortium }\href{http://paperpile.com/b/BKye4I/ovLNj}{\textit{et al.}\href{http://paperpile.com/b/BKye4I/ovLNj}{} Single-cell transcriptomics of 20 mouse organs creates a Tabula Muris. }\href{http://paperpile.com/b/BKye4I/ovLNj}{\textit{Nature}\href{http://paperpile.com/b/BKye4I/ovLNj}{} }\href{http://paperpile.com/b/BKye4I/ovLNj}{\textbf{562}\href{http://paperpile.com/b/BKye4I/ovLNj}{}, 367–372 (2018).}\par}
\end{FlushLeft}\par

\end{adjustwidth}

\begin{adjustwidth}{0.31in}{0.0in}
\begin{FlushLeft}
{\fontsize{11pt}{13.2pt}\selectfont 18.\tab \href{http://paperpile.com/b/BKye4I/1T9D1}{Vento-Tormo, R. }\href{http://paperpile.com/b/BKye4I/1T9D1}{\textit{et al.}\href{http://paperpile.com/b/BKye4I/1T9D1}{} Single-cell reconstruction of the early maternal-fetal interface in humans. }\href{http://paperpile.com/b/BKye4I/1T9D1}{\textit{Nature}\href{http://paperpile.com/b/BKye4I/1T9D1}{} }\href{http://paperpile.com/b/BKye4I/1T9D1}{\textbf{563}\href{http://paperpile.com/b/BKye4I/1T9D1}{}, 347–353 (2018).}\par}
\end{FlushLeft}\par

\end{adjustwidth}

\begin{adjustwidth}{0.31in}{0.0in}
\begin{FlushLeft}
{\fontsize{11pt}{13.2pt}\selectfont 19.\tab \href{http://paperpile.com/b/BKye4I/3ZMeu}{Vieira Braga, F. A. }\href{http://paperpile.com/b/BKye4I/3ZMeu}{\textit{et al.}\href{http://paperpile.com/b/BKye4I/3ZMeu}{} A cellular census of human lungs identifies novel cell states in health and in asthma. }\href{http://paperpile.com/b/BKye4I/3ZMeu}{\textit{Nat. Med.}\href{http://paperpile.com/b/BKye4I/3ZMeu}{} (2019). doi:}\href{http://dx.doi.org/10.1038/s41591-019-0468-5}{10.1038/s41591-019-0468-5}\par}
\end{FlushLeft}\par

\end{adjustwidth}

\begin{adjustwidth}{0.31in}{0.0in}
\begin{FlushLeft}
{\fontsize{11pt}{13.2pt}\selectfont 20.\tab \href{http://paperpile.com/b/BKye4I/4AUxR}{Kiselev, V. Y. }\href{http://paperpile.com/b/BKye4I/4AUxR}{\textit{et al.}\href{http://paperpile.com/b/BKye4I/4AUxR}{} SC3: consensus clustering of single-cell RNA-seq data. }\href{http://paperpile.com/b/BKye4I/4AUxR}{\textit{Nat. Methods}\href{http://paperpile.com/b/BKye4I/4AUxR}{} }\href{http://paperpile.com/b/BKye4I/4AUxR}{\textbf{14}\href{http://paperpile.com/b/BKye4I/4AUxR}{}, 483–486 (2017).}\par}
\end{FlushLeft}\par

\end{adjustwidth}

\begin{adjustwidth}{0.31in}{0.0in}
\begin{FlushLeft}
{\fontsize{11pt}{13.2pt}\selectfont 21.\tab \href{http://paperpile.com/b/BKye4I/NRIXy}{Zhang, J. M., Fan, J., Fan, H. C., Rosenfeld, D. $\&$  Tse, D. N. An interpretable framework for clustering single-cell RNA-Seq datasets. }\href{http://paperpile.com/b/BKye4I/NRIXy}{\textit{BMC Bioinformatics}\href{http://paperpile.com/b/BKye4I/NRIXy}{} }\href{http://paperpile.com/b/BKye4I/NRIXy}{\textbf{19}\href{http://paperpile.com/b/BKye4I/NRIXy}{}, 93 (2018).}\par}
\end{FlushLeft}\par

\end{adjustwidth}

\begin{adjustwidth}{0.31in}{0.0in}
\begin{FlushLeft}
{\fontsize{11pt}{13.2pt}\selectfont 22.\tab \href{http://paperpile.com/b/BKye4I/z6JA}{de Kanter, J. K., Lijnzaad, P., Candelli, T., Margaritis, T. $\&$  Holstege, F. C. P. CHETAH: a selective, hierarchical cell type identification method for single-cell RNA sequencing. }\href{http://paperpile.com/b/BKye4I/z6JA}{\textit{Nucleic Acids Res.}\href{http://paperpile.com/b/BKye4I/z6JA}{} (2019). doi:}\href{http://dx.doi.org/10.1093/nar/gkz543}{10.1093/nar/gkz543}\par}
\end{FlushLeft}\par

\end{adjustwidth}

\begin{adjustwidth}{0.31in}{0.0in}
\begin{FlushLeft}
{\fontsize{11pt}{13.2pt}\selectfont 23.\tab \href{http://paperpile.com/b/BKye4I/ia4B}{Xie, P. }\href{http://paperpile.com/b/BKye4I/ia4B}{\textit{et al.}\href{http://paperpile.com/b/BKye4I/ia4B}{} SuperCT: a supervised-learning framework for enhanced characterization of single-cell transcriptomic profiles. }\href{http://paperpile.com/b/BKye4I/ia4B}{\textit{Nucleic Acids Res.}\href{http://paperpile.com/b/BKye4I/ia4B}{} }\href{http://paperpile.com/b/BKye4I/ia4B}{\textbf{47}\href{http://paperpile.com/b/BKye4I/ia4B}{}, e48 (2019).}\par}
\end{FlushLeft}\par

\end{adjustwidth}

\begin{adjustwidth}{0.31in}{0.0in}
\begin{FlushLeft}
{\fontsize{11pt}{13.2pt}\selectfont 24.\tab \href{http://paperpile.com/b/BKye4I/mL7w}{Pliner, H. A., Shendure, J. $\&$  Trapnell, C. Supervised classification enables rapid annotation of cell atlases. }\href{http://paperpile.com/b/BKye4I/mL7w}{\textit{bioRxiv}\href{http://paperpile.com/b/BKye4I/mL7w}{} 538652 (2019). doi:}\href{http://dx.doi.org/10.1101/538652}{10.1101/538652}\par}
\end{FlushLeft}\par

\end{adjustwidth}

\begin{adjustwidth}{0.31in}{0.0in}
\begin{FlushLeft}
{\fontsize{11pt}{13.2pt}\selectfont 25.\tab \href{http://paperpile.com/b/BKye4I/cWTh}{Zhang, A. W. }\href{http://paperpile.com/b/BKye4I/cWTh}{\textit{et al.}\href{http://paperpile.com/b/BKye4I/cWTh}{} Probabilistic cell type assignment of single-cell transcriptomic data reveals spatiotemporal microenvironment dynamics in human cancers. }\href{http://paperpile.com/b/BKye4I/cWTh}{\textit{bioRxiv}\href{http://paperpile.com/b/BKye4I/cWTh}{} 521914 (2019). doi:}\href{http://dx.doi.org/10.1101/521914}{10.1101/521914}\par}
\end{FlushLeft}\par

\end{adjustwidth}

\begin{adjustwidth}{0.31in}{0.0in}
\begin{FlushLeft}
{\fontsize{11pt}{13.2pt}\selectfont 26.\tab \href{http://paperpile.com/b/BKye4I/YStx}{Aran, D. }\href{http://paperpile.com/b/BKye4I/YStx}{\textit{et al.}\href{http://paperpile.com/b/BKye4I/YStx}{} Reference-based analysis of lung single-cell sequencing reveals a transitional profibrotic macrophage. }\href{http://paperpile.com/b/BKye4I/YStx}{\textit{Nat. Immunol.}\href{http://paperpile.com/b/BKye4I/YStx}{} }\href{http://paperpile.com/b/BKye4I/YStx}{\textbf{20}\href{http://paperpile.com/b/BKye4I/YStx}{}, 163–172 (2019).}\par}
\end{FlushLeft}\par

\end{adjustwidth}

\begin{adjustwidth}{0.31in}{0.0in}
\begin{FlushLeft}
{\fontsize{11pt}{13.2pt}\selectfont 27.\tab \href{http://paperpile.com/b/BKye4I/1lv9}{Tan, Y. $\&$  Cahan, P. SingleCellNet: A Computational Tool to Classify Single Cell RNA-Seq Data Across Platforms and Across Species. }\href{http://paperpile.com/b/BKye4I/1lv9}{\textit{Cell Syst}\href{http://paperpile.com/b/BKye4I/1lv9}{} }\href{http://paperpile.com/b/BKye4I/1lv9}{\textbf{9}\href{http://paperpile.com/b/BKye4I/1lv9}{}, 207–213.e2 (2019).}\par}
\end{FlushLeft}\par

\end{adjustwidth}

\begin{adjustwidth}{0.31in}{0.0in}
\begin{FlushLeft}
{\fontsize{11pt}{13.2pt}\selectfont 28.\tab \href{http://paperpile.com/b/BKye4I/J2VQ}{Wagner, F. $\&$  Yanai, I. Moana: A robust and scalable cell type classification framework for single-cell RNA-Seq data. doi:}\href{http://dx.doi.org/10.1101/456129}{10.1101/456129}\par}
\end{FlushLeft}\par

\end{adjustwidth}

\begin{adjustwidth}{0.31in}{0.0in}
\begin{FlushLeft}
{\fontsize{11pt}{13.2pt}\selectfont 29.\tab \href{http://paperpile.com/b/BKye4I/zaRi}{Ma, F. $\&$  Pellegrini, M. ACTINN: Automated Identification of Cell Types in Single Cell RNA Sequencing. }\href{http://paperpile.com/b/BKye4I/zaRi}{\textit{Bioinformatics}\href{http://paperpile.com/b/BKye4I/zaRi}{} (2019). doi:}\href{http://dx.doi.org/10.1093/bioinformatics/btz592}{10.1093/bioinformatics/btz592}\par}
\end{FlushLeft}\par

\end{adjustwidth}

\begin{adjustwidth}{0.31in}{0.0in}
\begin{FlushLeft}
{\fontsize{11pt}{13.2pt}\selectfont 30.\tab \href{http://paperpile.com/b/BKye4I/kYlT}{Lin, Y. }\href{http://paperpile.com/b/BKye4I/kYlT}{\textit{et al.}\href{http://paperpile.com/b/BKye4I/kYlT}{} scClassify: hierarchical classification of cells. doi:}\href{http://dx.doi.org/10.1101/776948}{10.1101/776948}\par}
\end{FlushLeft}\par

\end{adjustwidth}

\begin{adjustwidth}{0.31in}{0.0in}
\begin{FlushLeft}
{\fontsize{11pt}{13.2pt}\selectfont 31.\tab \href{http://paperpile.com/b/BKye4I/var3u}{Ntranos, V., Yi, L., Melsted, P. $\&$  Pachter, L. A discriminative learning approach to differential expression analysis for single-cell RNA-seq. }\href{http://paperpile.com/b/BKye4I/var3u}{\textit{Nat. Methods}\href{http://paperpile.com/b/BKye4I/var3u}{} }\href{http://paperpile.com/b/BKye4I/var3u}{\textbf{16}\href{http://paperpile.com/b/BKye4I/var3u}{}, 163–166 (2019).}\par}
\end{FlushLeft}\par

\end{adjustwidth}

\begin{adjustwidth}{0.31in}{0.0in}
\begin{FlushLeft}
{\fontsize{11pt}{13.2pt}\selectfont 32.\tab \href{http://paperpile.com/b/BKye4I/fFIer}{Zappia, L., Phipson, B. $\&$  Oshlack, A. Splatter: simulation of single-cell RNA sequencing data. }\href{http://paperpile.com/b/BKye4I/fFIer}{\textit{Genome Biol.}\href{http://paperpile.com/b/BKye4I/fFIer}{} }\href{http://paperpile.com/b/BKye4I/fFIer}{\textbf{18}\href{http://paperpile.com/b/BKye4I/fFIer}{}, 174 (2017).}\par}
\end{FlushLeft}\par

\end{adjustwidth}

\begin{adjustwidth}{0.31in}{0.0in}
\begin{FlushLeft}
{\fontsize{11pt}{13.2pt}\selectfont 33.\tab \href{http://paperpile.com/b/BKye4I/Tsuir}{Dimitriadis, G., Neto, J. P. $\&$  Kampff, A. R. t-SNE Visualization of Large-Scale Neural Recordings. }\href{http://paperpile.com/b/BKye4I/Tsuir}{\textit{Neural Comput.}\href{http://paperpile.com/b/BKye4I/Tsuir}{} }\href{http://paperpile.com/b/BKye4I/Tsuir}{\textbf{30}\href{http://paperpile.com/b/BKye4I/Tsuir}{}, 1750–1774 (2018).}\par}
\end{FlushLeft}\par

\end{adjustwidth}

\begin{adjustwidth}{0.31in}{0.0in}
\begin{FlushLeft}
{\fontsize{11pt}{13.2pt}\selectfont 34.\tab \href{http://paperpile.com/b/BKye4I/cbqJ}{Hubert, L. $\&$  Arabie, P. Comparing partitions. }\href{http://paperpile.com/b/BKye4I/cbqJ}{\textit{J. Classification}\href{http://paperpile.com/b/BKye4I/cbqJ}{} }\href{http://paperpile.com/b/BKye4I/cbqJ}{\textbf{2}\href{http://paperpile.com/b/BKye4I/cbqJ}{}, 193–218 (1985).}\par}
\end{FlushLeft}\par

\end{adjustwidth}

\begin{adjustwidth}{0.31in}{0.0in}
\begin{FlushLeft}
{\fontsize{11pt}{13.2pt}\selectfont 35.\tab \href{http://paperpile.com/b/BKye4I/Gfzwj}{Zeisel, A. }\href{http://paperpile.com/b/BKye4I/Gfzwj}{\textit{et al.}\href{http://paperpile.com/b/BKye4I/Gfzwj}{} Brain structure. Cell types in the mouse cortex and hippocampus revealed by single-cell RNA-seq. }\href{http://paperpile.com/b/BKye4I/Gfzwj}{\textit{Science}\href{http://paperpile.com/b/BKye4I/Gfzwj}{} }\href{http://paperpile.com/b/BKye4I/Gfzwj}{\textbf{347}\href{http://paperpile.com/b/BKye4I/Gfzwj}{}, 1138–1142 (2015).}\par}
\end{FlushLeft}\par

\end{adjustwidth}

\begin{adjustwidth}{0.31in}{0.0in}
\begin{FlushLeft}
{\fontsize{11pt}{13.2pt}\selectfont 36.\tab \href{http://paperpile.com/b/BKye4I/mJd6E}{Segerstolpe, Å. }\href{http://paperpile.com/b/BKye4I/mJd6E}{\textit{et al.}\href{http://paperpile.com/b/BKye4I/mJd6E}{} Single-Cell Transcriptome Profiling of Human Pancreatic Islets in Health and Type 2 Diabetes. }\href{http://paperpile.com/b/BKye4I/mJd6E}{\textit{Cell Metab.}\href{http://paperpile.com/b/BKye4I/mJd6E}{} }\href{http://paperpile.com/b/BKye4I/mJd6E}{\textbf{24}\href{http://paperpile.com/b/BKye4I/mJd6E}{}, 593–607 (2016).}\par}
\end{FlushLeft}\par

\end{adjustwidth}

\begin{adjustwidth}{0.31in}{0.0in}
\begin{FlushLeft}
{\fontsize{11pt}{13.2pt}\selectfont 37.\tab \href{http://paperpile.com/b/BKye4I/eE5EA}{Hrvatin, S. }\href{http://paperpile.com/b/BKye4I/eE5EA}{\textit{et al.}\href{http://paperpile.com/b/BKye4I/eE5EA}{} Single-cell analysis of experience-dependent transcriptomic states in the mouse visual cortex. }\href{http://paperpile.com/b/BKye4I/eE5EA}{\textit{Nat. Neurosci.}\href{http://paperpile.com/b/BKye4I/eE5EA}{} }\href{http://paperpile.com/b/BKye4I/eE5EA}{\textbf{21}\href{http://paperpile.com/b/BKye4I/eE5EA}{}, 120–129 (2018).}\par}
\end{FlushLeft}\par

\end{adjustwidth}

\begin{adjustwidth}{0.31in}{0.0in}
\begin{FlushLeft}
{\fontsize{11pt}{13.2pt}\selectfont 38.\tab \href{http://paperpile.com/b/BKye4I/otnx}{Aizarani, N. }\href{http://paperpile.com/b/BKye4I/otnx}{\textit{et al.}\href{http://paperpile.com/b/BKye4I/otnx}{} A human liver cell atlas reveals heterogeneity and epithelial progenitors. }\href{http://paperpile.com/b/BKye4I/otnx}{\textit{Nature}\href{http://paperpile.com/b/BKye4I/otnx}{} }\href{http://paperpile.com/b/BKye4I/otnx}{\textbf{572}\href{http://paperpile.com/b/BKye4I/otnx}{}, 199–204 (2019).}\par}
\end{FlushLeft}\par

\end{adjustwidth}

\begin{adjustwidth}{0.31in}{0.0in}
\begin{FlushLeft}
{\fontsize{11pt}{13.2pt}\selectfont 39.\tab \href{http://paperpile.com/b/BKye4I/AXkh}{Tasic, B. }\href{http://paperpile.com/b/BKye4I/AXkh}{\textit{et al.}\href{http://paperpile.com/b/BKye4I/AXkh}{} Shared and distinct transcriptomic cell types across neocortical areas. }\href{http://paperpile.com/b/BKye4I/AXkh}{\textit{Nature}\href{http://paperpile.com/b/BKye4I/AXkh}{} }\href{http://paperpile.com/b/BKye4I/AXkh}{\textbf{563}\href{http://paperpile.com/b/BKye4I/AXkh}{}, 72–78 (2018).}\par}
\end{FlushLeft}\par

\end{adjustwidth}

\begin{adjustwidth}{0.31in}{0.0in}
\begin{FlushLeft}
{\fontsize{11pt}{13.2pt}\selectfont 40.\tab \href{http://paperpile.com/b/BKye4I/CeZ6x}{Tracy, C. A. $\&$  Widom, H. Level-spacing distributions and the Airy kernel. }\href{http://paperpile.com/b/BKye4I/CeZ6x}{\textit{Communications in Mathematical Physics}\href{http://paperpile.com/b/BKye4I/CeZ6x}{} }\href{http://paperpile.com/b/BKye4I/CeZ6x}{\textbf{159}\href{http://paperpile.com/b/BKye4I/CeZ6x}{}, 151–174 (1994).}\par}
\end{FlushLeft}\par

\end{adjustwidth}

\begin{adjustwidth}{0.31in}{0.0in}
\begin{FlushLeft}
{\fontsize{11pt}{13.2pt}\selectfont 41.\tab \href{http://paperpile.com/b/BKye4I/RdCh9}{Tusi, B. K. }\href{http://paperpile.com/b/BKye4I/RdCh9}{\textit{et al.}\href{http://paperpile.com/b/BKye4I/RdCh9}{} Population snapshots predict early haematopoietic and erythroid hierarchies. }\href{http://paperpile.com/b/BKye4I/RdCh9}{\textit{Nature}\href{http://paperpile.com/b/BKye4I/RdCh9}{} }\href{http://paperpile.com/b/BKye4I/RdCh9}{\textbf{555}\href{http://paperpile.com/b/BKye4I/RdCh9}{}, 54–60 (2018).}\par}
\end{FlushLeft}\par

\end{adjustwidth}

\begin{adjustwidth}{0.31in}{0.0in}
\begin{FlushLeft}
{\fontsize{11pt}{13.2pt}\selectfont 42.\tab \href{http://paperpile.com/b/BKye4I/DmgXV}{Giladi, A. }\href{http://paperpile.com/b/BKye4I/DmgXV}{\textit{et al.}\href{http://paperpile.com/b/BKye4I/DmgXV}{} Single-cell characterization of haematopoietic progenitors and their trajectories in homeostasis and perturbed haematopoiesis. }\href{http://paperpile.com/b/BKye4I/DmgXV}{\textit{Nat. Cell Biol.}\href{http://paperpile.com/b/BKye4I/DmgXV}{} }\href{http://paperpile.com/b/BKye4I/DmgXV}{\textbf{20}\href{http://paperpile.com/b/BKye4I/DmgXV}{}, 836–846 (2018).}\par}
\end{FlushLeft}\par

\end{adjustwidth}

\begin{adjustwidth}{0.31in}{0.0in}
\begin{FlushLeft}
{\fontsize{11pt}{13.2pt}\selectfont 43.\tab \href{http://paperpile.com/b/BKye4I/cgeDc}{McInnes, L. $\&$  Healy, J. UMAP: Uniform Manifold Approximation and Projection for Dimension Reduction. (2018).}\par}
\end{FlushLeft}\par

\end{adjustwidth}

\begin{adjustwidth}{0.31in}{0.0in}
\begin{FlushLeft}
{\fontsize{11pt}{13.2pt}\selectfont 44.\tab \href{http://paperpile.com/b/BKye4I/fFdsG}{Becht, E. }\href{http://paperpile.com/b/BKye4I/fFdsG}{\textit{et al.}\href{http://paperpile.com/b/BKye4I/fFdsG}{} Dimensionality reduction for visualizing single-cell data using UMAP. }\href{http://paperpile.com/b/BKye4I/fFdsG}{\textit{Nat. Biotechnol.}\href{http://paperpile.com/b/BKye4I/fFdsG}{} (2018). doi:}\href{http://dx.doi.org/10.1038/nbt.4314}{10.1038/nbt.4314}\par}
\end{FlushLeft}\par

\end{adjustwidth}

\begin{adjustwidth}{0.31in}{0.0in}
\begin{FlushLeft}
{\fontsize{11pt}{13.2pt}\selectfont 45.\tab \href{http://paperpile.com/b/BKye4I/u71mp}{Lee, J., Krivega, I., Dale, R. K. $\&$  Dean, A. The LDB1 Complex Co-opts CTCF for Erythroid Lineage-Specific Long-Range Enhancer Interactions. }\href{http://paperpile.com/b/BKye4I/u71mp}{\textit{Cell Rep.}\href{http://paperpile.com/b/BKye4I/u71mp}{} }\href{http://paperpile.com/b/BKye4I/u71mp}{\textbf{19}\href{http://paperpile.com/b/BKye4I/u71mp}{}, 2490–2502 (2017).}\par}
\end{FlushLeft}\par

\end{adjustwidth}

\begin{adjustwidth}{0.31in}{0.0in}
\begin{FlushLeft}
{\fontsize{11pt}{13.2pt}\selectfont 46.\tab \href{http://paperpile.com/b/BKye4I/n4P9o}{Song, S.-H., Kim, A., Dale, R. $\&$  Dean, A. Ldb1 regulates carbonic anhydrase 1 during erythroid differentiation. }\href{http://paperpile.com/b/BKye4I/n4P9o}{\textit{Biochim. Biophys. Acta}\href{http://paperpile.com/b/BKye4I/n4P9o}{} }\href{http://paperpile.com/b/BKye4I/n4P9o}{\textbf{1819}\href{http://paperpile.com/b/BKye4I/n4P9o}{}, 885–891 (2012).}\par}
\end{FlushLeft}\par

\end{adjustwidth}

\begin{adjustwidth}{0.31in}{0.0in}
\begin{FlushLeft}
{\fontsize{11pt}{13.2pt}\selectfont 47.\tab \href{http://paperpile.com/b/BKye4I/PvEjT}{Nemeth, M. J., Cline, A. P., Anderson, S. M., Garrett-Beal, L. J. $\&$  Bodine, D. M. Hmgb3 deficiency deregulates proliferation and differentiation of common lymphoid and myeloid progenitors. }\href{http://paperpile.com/b/BKye4I/PvEjT}{\textit{Blood}\href{http://paperpile.com/b/BKye4I/PvEjT}{} }\href{http://paperpile.com/b/BKye4I/PvEjT}{\textbf{105}\href{http://paperpile.com/b/BKye4I/PvEjT}{}, 627–634 (2005).}\par}
\end{FlushLeft}\par

\end{adjustwidth}

\begin{adjustwidth}{0.31in}{0.0in}
\begin{FlushLeft}
{\fontsize{11pt}{13.2pt}\selectfont 48.\tab \href{http://paperpile.com/b/BKye4I/84v9W}{Chi, H. }\href{http://paperpile.com/b/BKye4I/84v9W}{\textit{et al.}\href{http://paperpile.com/b/BKye4I/84v9W}{} Targeted deletion of Minpp1 provides new insight into the activity of multiple inositol polyphosphate phosphatase in vivo. }\href{http://paperpile.com/b/BKye4I/84v9W}{\textit{Mol. Cell. Biol.}\href{http://paperpile.com/b/BKye4I/84v9W}{} }\href{http://paperpile.com/b/BKye4I/84v9W}{\textbf{20}\href{http://paperpile.com/b/BKye4I/84v9W}{}, 6496–6507 (2000).}\par}
\end{FlushLeft}\par

\end{adjustwidth}

\begin{adjustwidth}{0.31in}{0.0in}
\begin{FlushLeft}
{\fontsize{11pt}{13.2pt}\selectfont 49.\tab \href{http://paperpile.com/b/BKye4I/7Oc5L}{Xiong, Y., Yang, P., Proia, R. L. $\&$  Hla, T. Erythrocyte-derived sphingosine 1-phosphate is essential for vascular development. }\href{http://paperpile.com/b/BKye4I/7Oc5L}{\textit{J. Clin. Invest.}\href{http://paperpile.com/b/BKye4I/7Oc5L}{} }\href{http://paperpile.com/b/BKye4I/7Oc5L}{\textbf{124}\href{http://paperpile.com/b/BKye4I/7Oc5L}{}, 4823–4828 (2014).}\par}
\end{FlushLeft}\par

\end{adjustwidth}

\begin{adjustwidth}{0.31in}{0.0in}
\begin{FlushLeft}
{\fontsize{11pt}{13.2pt}\selectfont 50.\tab \href{http://paperpile.com/b/BKye4I/71Jlg}{Nemeth, M. J., Kirby, M. R. $\&$  Bodine, D. M. Hmgb3 regulates the balance between hematopoietic stem cell self-renewal and differentiation. }\href{http://paperpile.com/b/BKye4I/71Jlg}{\textit{Proc. Natl. Acad. Sci. U. S. A.}\href{http://paperpile.com/b/BKye4I/71Jlg}{} }\href{http://paperpile.com/b/BKye4I/71Jlg}{\textbf{103}\href{http://paperpile.com/b/BKye4I/71Jlg}{}, 13783–13788 (2006).}\par}
\end{FlushLeft}\par

\end{adjustwidth}

\begin{adjustwidth}{0.31in}{0.0in}
\begin{FlushLeft}
{\fontsize{11pt}{13.2pt}\selectfont 51.\tab \href{http://paperpile.com/b/BKye4I/SEtgs}{Forster, L. }\href{http://paperpile.com/b/BKye4I/SEtgs}{\textit{et al.}\href{http://paperpile.com/b/BKye4I/SEtgs}{} Differential gene expression analysis in early and late erythroid progenitor cells in $ \beta $ -thalassaemia. }\href{http://paperpile.com/b/BKye4I/SEtgs}{\textit{British Journal of Haematology}\href{http://paperpile.com/b/BKye4I/SEtgs}{} }\href{http://paperpile.com/b/BKye4I/SEtgs}{\textbf{170}\href{http://paperpile.com/b/BKye4I/SEtgs}{}, 257–267 (2015).}\par}
\end{FlushLeft}\par

\end{adjustwidth}

\begin{adjustwidth}{0.31in}{0.0in}
\begin{FlushLeft}
{\fontsize{11pt}{13.2pt}\selectfont 52.\tab \href{http://paperpile.com/b/BKye4I/rhD10}{Yang, C., Hashimoto, M., Lin, Q. X. X., Tan, D. Q. $\&$  Suda, T. Sphingosine-1-phosphate signaling modulates terminal erythroid differentiation through the regulation of mitophagy. }\href{http://paperpile.com/b/BKye4I/rhD10}{\textit{Exp. Hematol.}\href{http://paperpile.com/b/BKye4I/rhD10}{} }\href{http://paperpile.com/b/BKye4I/rhD10}{\textbf{72}\href{http://paperpile.com/b/BKye4I/rhD10}{}, 47–59.e1 (2019).}\par}
\end{FlushLeft}\par

\end{adjustwidth}

\begin{adjustwidth}{0.31in}{0.0in}
\begin{FlushLeft}
{\fontsize{11pt}{13.2pt}\selectfont 53.\tab \href{http://paperpile.com/b/BKye4I/Jzx3W}{Kim, Y.-H. }\href{http://paperpile.com/b/BKye4I/Jzx3W}{\textit{et al.}\href{http://paperpile.com/b/BKye4I/Jzx3W}{} Rh D blood group conversion using transcription activator-like effector nucleases. }\href{http://paperpile.com/b/BKye4I/Jzx3W}{\textit{Nat. Commun.}\href{http://paperpile.com/b/BKye4I/Jzx3W}{} }\href{http://paperpile.com/b/BKye4I/Jzx3W}{\textbf{6}\href{http://paperpile.com/b/BKye4I/Jzx3W}{}, 7451 (2015).}\par}
\end{FlushLeft}\par

\end{adjustwidth}

\begin{adjustwidth}{0.31in}{0.0in}
\begin{FlushLeft}
{\fontsize{11pt}{13.2pt}\selectfont 54.\tab \href{http://paperpile.com/b/BKye4I/mnK7O}{Singh, R. P. }\href{http://paperpile.com/b/BKye4I/mnK7O}{\textit{et al.}\href{http://paperpile.com/b/BKye4I/mnK7O}{} Hematopoietic Stem Cells but Not Multipotent Progenitors Drive Erythropoiesis during Chronic Erythroid Stress in EPO Transgenic Mice. }\href{http://paperpile.com/b/BKye4I/mnK7O}{\textit{Stem Cell Reports}\href{http://paperpile.com/b/BKye4I/mnK7O}{} }\href{http://paperpile.com/b/BKye4I/mnK7O}{\textbf{10}\href{http://paperpile.com/b/BKye4I/mnK7O}{}, 1908–1919 (2018).}\par}
\end{FlushLeft}\par

\end{adjustwidth}

\begin{adjustwidth}{0.31in}{0.0in}
\begin{FlushLeft}
{\fontsize{11pt}{13.2pt}\selectfont 55.\tab \href{http://paperpile.com/b/BKye4I/Xx5mf}{Subramanian, G. }\href{http://paperpile.com/b/BKye4I/Xx5mf}{\textit{et al.}\href{http://paperpile.com/b/BKye4I/Xx5mf}{} Lamin B receptor regulates the growth and maturation of myeloid progenitors via its sterol reductase domain: implications for cholesterol biosynthesis in regulating myelopoiesis. }\href{http://paperpile.com/b/BKye4I/Xx5mf}{\textit{J. Immunol.}\href{http://paperpile.com/b/BKye4I/Xx5mf}{} }\href{http://paperpile.com/b/BKye4I/Xx5mf}{\textbf{188}\href{http://paperpile.com/b/BKye4I/Xx5mf}{}, 85–102 (2012).}\par}
\end{FlushLeft}\par

\end{adjustwidth}

\begin{adjustwidth}{0.31in}{0.0in}
\begin{FlushLeft}
{\fontsize{11pt}{13.2pt}\selectfont 56.\tab \href{http://paperpile.com/b/BKye4I/JDu3G}{McIver, S. C. }\href{http://paperpile.com/b/BKye4I/JDu3G}{\textit{et al.}\href{http://paperpile.com/b/BKye4I/JDu3G}{} The exosome complex establishes a barricade to erythroid maturation. }\href{http://paperpile.com/b/BKye4I/JDu3G}{\textit{Blood}\href{http://paperpile.com/b/BKye4I/JDu3G}{} }\href{http://paperpile.com/b/BKye4I/JDu3G}{\textbf{124}\href{http://paperpile.com/b/BKye4I/JDu3G}{}, 2285–2297 (2014).}\par}
\end{FlushLeft}\par

\end{adjustwidth}

\begin{adjustwidth}{0.31in}{0.0in}
\begin{FlushLeft}
{\fontsize{11pt}{13.2pt}\selectfont 57.\tab \href{http://paperpile.com/b/BKye4I/y5MHL}{Leung, C. G. }\href{http://paperpile.com/b/BKye4I/y5MHL}{\textit{et al.}\href{http://paperpile.com/b/BKye4I/y5MHL}{} Requirements for survivin in terminal differentiation of erythroid cells and maintenance of hematopoietic stem and progenitor cells. }\href{http://paperpile.com/b/BKye4I/y5MHL}{\textit{J. Exp. Med.}\href{http://paperpile.com/b/BKye4I/y5MHL}{} }\href{http://paperpile.com/b/BKye4I/y5MHL}{\textbf{204}\href{http://paperpile.com/b/BKye4I/y5MHL}{}, 1603–1611 (2007).}\par}
\end{FlushLeft}\par

\end{adjustwidth}

\begin{adjustwidth}{0.31in}{0.0in}
\begin{FlushLeft}
{\fontsize{11pt}{13.2pt}\selectfont 58.\tab \href{http://paperpile.com/b/BKye4I/yQQHY}{Kingsley, P. D. }\href{http://paperpile.com/b/BKye4I/yQQHY}{\textit{et al.}\href{http://paperpile.com/b/BKye4I/yQQHY}{} ‘Maturational’ globin switching in primary primitive erythroid cells. }\href{http://paperpile.com/b/BKye4I/yQQHY}{\textit{Blood}\href{http://paperpile.com/b/BKye4I/yQQHY}{} }\href{http://paperpile.com/b/BKye4I/yQQHY}{\textbf{107}\href{http://paperpile.com/b/BKye4I/yQQHY}{}, 1665–1672 (2006).}\par}
\end{FlushLeft}\par

\end{adjustwidth}

\begin{adjustwidth}{0.31in}{0.0in}
\begin{FlushLeft}
{\fontsize{11pt}{13.2pt}\selectfont 59.\tab \href{http://paperpile.com/b/BKye4I/C2j9K}{Petousi, N. }\href{http://paperpile.com/b/BKye4I/C2j9K}{\textit{et al.}\href{http://paperpile.com/b/BKye4I/C2j9K}{} Erythrocytosis associated with a novel missense mutation in the BPGM gene. }\href{http://paperpile.com/b/BKye4I/C2j9K}{\textit{Haematologica}\href{http://paperpile.com/b/BKye4I/C2j9K}{} }\href{http://paperpile.com/b/BKye4I/C2j9K}{\textbf{99}\href{http://paperpile.com/b/BKye4I/C2j9K}{}, e201–4 (2014).}\par}
\end{FlushLeft}\par

\end{adjustwidth}

\begin{adjustwidth}{0.31in}{0.0in}
\begin{FlushLeft}
{\fontsize{11pt}{13.2pt}\selectfont 60.\tab \href{http://paperpile.com/b/BKye4I/0lrwd}{Huang, Y. }\href{http://paperpile.com/b/BKye4I/0lrwd}{\textit{et al.}\href{http://paperpile.com/b/BKye4I/0lrwd}{} SF3B1 deficiency impairs human erythropoiesis via activation of p53 pathway: implications for understanding of ineffective erythropoiesis in MDS. }\href{http://paperpile.com/b/BKye4I/0lrwd}{\textit{J. Hematol. Oncol.}\href{http://paperpile.com/b/BKye4I/0lrwd}{} }\href{http://paperpile.com/b/BKye4I/0lrwd}{\textbf{11}\href{http://paperpile.com/b/BKye4I/0lrwd}{}, 19 (2018).}\par}
\end{FlushLeft}\par

\end{adjustwidth}

\begin{adjustwidth}{0.31in}{0.0in}
\begin{FlushLeft}
{\fontsize{11pt}{13.2pt}\selectfont 61.\tab \href{http://paperpile.com/b/BKye4I/Ix94B}{Liang, R. }\href{http://paperpile.com/b/BKye4I/Ix94B}{\textit{et al.}\href{http://paperpile.com/b/BKye4I/Ix94B}{} A Systems Approach Identifies Essential FOXO3 Functions at Key Steps of Terminal Erythropoiesis. }\href{http://paperpile.com/b/BKye4I/Ix94B}{\textit{PLoS Genet.}\href{http://paperpile.com/b/BKye4I/Ix94B}{} }\href{http://paperpile.com/b/BKye4I/Ix94B}{\textbf{11}\href{http://paperpile.com/b/BKye4I/Ix94B}{}, e1005526 (2015).}\par}
\end{FlushLeft}\par

\end{adjustwidth}

\begin{adjustwidth}{0.31in}{0.0in}
\begin{FlushLeft}
{\fontsize{11pt}{13.2pt}\selectfont 62.\tab \href{http://paperpile.com/b/BKye4I/CJEmu}{Sadlon, T. J., Dell’Oso, T., Surinya, K. H. $\&$  May, B. K. Regulation of erythroid 5-aminolevulinate synthase expression during erythropoiesis. }\href{http://paperpile.com/b/BKye4I/CJEmu}{\textit{Int. J. Biochem. Cell Biol.}\href{http://paperpile.com/b/BKye4I/CJEmu}{} }\href{http://paperpile.com/b/BKye4I/CJEmu}{\textbf{31}\href{http://paperpile.com/b/BKye4I/CJEmu}{}, 1153–1167 (1999).}\par}
\end{FlushLeft}\par

\end{adjustwidth}

\begin{adjustwidth}{0.31in}{0.0in}
\begin{FlushLeft}
{\fontsize{11pt}{13.2pt}\selectfont 63.\tab \href{http://paperpile.com/b/BKye4I/2CQh}{Konstantinides, N. }\href{http://paperpile.com/b/BKye4I/2CQh}{\textit{et al.}\href{http://paperpile.com/b/BKye4I/2CQh}{} Phenotypic Convergence: Distinct Transcription Factors Regulate Common Terminal Features. }\href{http://paperpile.com/b/BKye4I/2CQh}{\textit{Cell}\href{http://paperpile.com/b/BKye4I/2CQh}{} }\href{http://paperpile.com/b/BKye4I/2CQh}{\textbf{174}\href{http://paperpile.com/b/BKye4I/2CQh}{}, 622–635.e13 (2018).}\par}
\end{FlushLeft}\par

\end{adjustwidth}

\begin{adjustwidth}{0.31in}{0.0in}
\begin{FlushLeft}
{\fontsize{11pt}{13.2pt}\selectfont 64.\tab \href{http://paperpile.com/b/BKye4I/3Ubg}{Gerber, T. }\href{http://paperpile.com/b/BKye4I/3Ubg}{\textit{et al.}\href{http://paperpile.com/b/BKye4I/3Ubg}{} Single-cell analysis uncovers convergence of cell identities during axolotl limb regeneration. }\href{http://paperpile.com/b/BKye4I/3Ubg}{\textit{Science}\href{http://paperpile.com/b/BKye4I/3Ubg}{} }\href{http://paperpile.com/b/BKye4I/3Ubg}{\textbf{362}\href{http://paperpile.com/b/BKye4I/3Ubg}{}, (2018).}\par}
\end{FlushLeft}\par

\end{adjustwidth}

\begin{adjustwidth}{0.31in}{0.0in}
\begin{FlushLeft}
{\fontsize{11pt}{13.2pt}\selectfont 65.\tab \href{http://paperpile.com/b/BKye4I/0WGt}{Afgan, E. }\href{http://paperpile.com/b/BKye4I/0WGt}{\textit{et al.}\href{http://paperpile.com/b/BKye4I/0WGt}{} The Galaxy platform for accessible, reproducible and collaborative biomedical analyses: 2018 update. }\href{http://paperpile.com/b/BKye4I/0WGt}{\textit{Nucleic Acids Res.}\href{http://paperpile.com/b/BKye4I/0WGt}{} }\href{http://paperpile.com/b/BKye4I/0WGt}{\textbf{46}\href{http://paperpile.com/b/BKye4I/0WGt}{}, W537–W544 (2018).}\par}
\end{FlushLeft}\par

\end{adjustwidth}

\begin{adjustwidth}{0.31in}{0.0in}
\begin{FlushLeft}
{\fontsize{11pt}{13.2pt}\selectfont 66.\tab \href{http://paperpile.com/b/BKye4I/gZUmc}{Nelli, F. Machine Learning with scikit-learn. in }\href{http://paperpile.com/b/BKye4I/gZUmc}{\textit{Python Data Analytics}\href{http://paperpile.com/b/BKye4I/gZUmc}{} 237–264 (2015).}\par}
\end{FlushLeft}\par

\end{adjustwidth}

\begin{adjustwidth}{0.31in}{0.0in}
\begin{FlushLeft}
{\fontsize{11pt}{13.2pt}\selectfont 67.\tab \href{http://paperpile.com/b/BKye4I/7oPv}{Stehman, S. V. Selecting and interpreting measures of thematic classification accuracy. }\href{http://paperpile.com/b/BKye4I/7oPv}{\textit{Remote Sensing of Environment}\href{http://paperpile.com/b/BKye4I/7oPv}{} }\href{http://paperpile.com/b/BKye4I/7oPv}{\textbf{62}\href{http://paperpile.com/b/BKye4I/7oPv}{}, 77–89 (1997).}\par}
\end{FlushLeft}\par

\end{adjustwidth}

\begin{adjustwidth}{0.31in}{0.0in}
\begin{FlushLeft}
{\fontsize{11pt}{13.2pt}\selectfont 68.\tab \href{http://paperpile.com/b/BKye4I/gfgWA}{Hunter, J. D. Matplotlib: A 2D Graphics Environment. }\href{http://paperpile.com/b/BKye4I/gfgWA}{\textit{Comput. Sci. Eng.}\href{http://paperpile.com/b/BKye4I/gfgWA}{} }\href{http://paperpile.com/b/BKye4I/gfgWA}{\textbf{9}\href{http://paperpile.com/b/BKye4I/gfgWA}{}, 90–95 (2007).}\par}
\end{FlushLeft}\par

\end{adjustwidth}

\begin{adjustwidth}{0.31in}{0.0in}
\begin{FlushLeft}
{\fontsize{11pt}{13.2pt}\selectfont 69.\tab \href{http://paperpile.com/b/BKye4I/DWXyj}{Hill, C. SciPy. in }\href{http://paperpile.com/b/BKye4I/DWXyj}{\textit{Learning Scientific Programming with Python}\href{http://paperpile.com/b/BKye4I/DWXyj}{} 333–401}\par}
\end{FlushLeft}\par

\end{adjustwidth}

\begin{adjustwidth}{0.31in}{0.0in}
\begin{FlushLeft}
{\fontsize{11pt}{13.2pt}\selectfont 70.\tab \href{http://paperpile.com/b/BKye4I/ZLP4X}{Svensson, V., Teichmann, S. A. $\&$  Stegle, O. SpatialDE: identification of spatially variable genes. }\href{http://paperpile.com/b/BKye4I/ZLP4X}{\textit{Nat. Methods}\href{http://paperpile.com/b/BKye4I/ZLP4X}{} }\href{http://paperpile.com/b/BKye4I/ZLP4X}{\textbf{15}\href{http://paperpile.com/b/BKye4I/ZLP4X}{}, 343–346 (2018).}\par}
\end{FlushLeft}\par

\end{adjustwidth}

\begin{adjustwidth}{0.31in}{0.0in}
\begin{FlushLeft}
{\fontsize{11pt}{13.2pt}\selectfont 71.\tab \href{http://paperpile.com/b/BKye4I/QNup}{Tasic, B. }\href{http://paperpile.com/b/BKye4I/QNup}{\textit{et al.}\href{http://paperpile.com/b/BKye4I/QNup}{} Adult mouse cortical cell taxonomy revealed by single cell transcriptomics. }\href{http://paperpile.com/b/BKye4I/QNup}{\textit{Nature Neuroscience}\href{http://paperpile.com/b/BKye4I/QNup}{} }\href{http://paperpile.com/b/BKye4I/QNup}{\textbf{19}\href{http://paperpile.com/b/BKye4I/QNup}{}, 335–346 (2016).}\par}
\end{FlushLeft}\par

\end{adjustwidth}

\begin{adjustwidth}{0.31in}{0.0in}
\begin{FlushLeft}
{\fontsize{11pt}{13.2pt}\selectfont 72.\tab \href{http://paperpile.com/b/BKye4I/ea2W}{MacParland, S. A. }\href{http://paperpile.com/b/BKye4I/ea2W}{\textit{et al.}\href{http://paperpile.com/b/BKye4I/ea2W}{} Single cell RNA sequencing of human liver reveals distinct intrahepatic macrophage populations. }\href{http://paperpile.com/b/BKye4I/ea2W}{\textit{Nature Communications}\href{http://paperpile.com/b/BKye4I/ea2W}{} }\href{http://paperpile.com/b/BKye4I/ea2W}{\textbf{9}\href{http://paperpile.com/b/BKye4I/ea2W}{}, (2018).}\par}
\end{FlushLeft}\par

\end{adjustwidth}

\begin{adjustwidth}{0.31in}{0.0in}
\begin{FlushLeft}
{\fontsize{11pt}{13.2pt}\selectfont 73.\tab \href{http://paperpile.com/b/BKye4I/qMDKz}{Leek, J. T. svaseq: removing batch effects and other unwanted noise from sequencing data. }\href{http://paperpile.com/b/BKye4I/qMDKz}{\textit{Nucleic Acids Res.}\href{http://paperpile.com/b/BKye4I/qMDKz}{} }\href{http://paperpile.com/b/BKye4I/qMDKz}{\textbf{42}\href{http://paperpile.com/b/BKye4I/qMDKz}{}, (2014).}\par}
\end{FlushLeft}\par

\end{adjustwidth}

\begin{adjustwidth}{0.31in}{0.0in}
\begin{FlushLeft}
{\fontsize{11pt}{13.2pt}\selectfont 74.\tab \href{http://paperpile.com/b/BKye4I/Bs9JL}{Azizi, E., Prabhakaran, S., Carr, A. $\&$  Pe’er, D. Bayesian Inference for Single-cell Clustering and Imputing. }\href{http://paperpile.com/b/BKye4I/Bs9JL}{\textit{Genomics and Computational Biology}\href{http://paperpile.com/b/BKye4I/Bs9JL}{} }\href{http://paperpile.com/b/BKye4I/Bs9JL}{\textbf{3}\href{http://paperpile.com/b/BKye4I/Bs9JL}{}, 46 (2017).}\par}
\end{FlushLeft}\par

\end{adjustwidth}

\begin{adjustwidth}{0.31in}{0.0in}
\begin{FlushLeft}
{\fontsize{11pt}{13.2pt}\selectfont 75.\tab \href{http://paperpile.com/b/BKye4I/nZodk}{Athar, A. }\href{http://paperpile.com/b/BKye4I/nZodk}{\textit{et al.}\href{http://paperpile.com/b/BKye4I/nZodk}{} ArrayExpress update - from bulk to single-cell expression data. }\href{http://paperpile.com/b/BKye4I/nZodk}{\textit{Nucleic Acids Res.}\href{http://paperpile.com/b/BKye4I/nZodk}{} }\href{http://paperpile.com/b/BKye4I/nZodk}{\textbf{47}\href{http://paperpile.com/b/BKye4I/nZodk}{}, D711–D715 (2019).}\par}
\end{FlushLeft}\par

\end{adjustwidth}

\setlength{\parskip}{11.04pt}
\begin{adjustwidth}{0.31in}{0.0in}
\begin{FlushLeft}
{\fontsize{11pt}{13.2pt}\selectfont 76.\tab \href{http://paperpile.com/b/BKye4I/oyhFn}{Allen Institute for Brain Science. Allen Cell Types Database. (2015). Available at: }\href{http://celltypes.brain-map.org.}{http://celltypes.brain-map.org.}\par}
\end{FlushLeft}\par

\end{adjustwidth}

\vspace{\baselineskip}
\vspace{\baselineskip}
\subsection*{Figure 1. A self-projection approach}
\addcontentsline{toc}{subsection}{Figure 1. A self-projection approach}
\begin{justify}
A scheme of machine learning based self-projection is shown in (\textbf{a}): i) randomly split the data into training and test set; ii) use the assigned clusters to train a machine learning classifier with cross-validation on the training set; iii) apply the machine learning model to the test set, which is considered as ‘self-projection’; iv) the consistency between the original cluster assignment of the test set and the self-projection result measures the reliability of the clustering.
\end{justify}\par

\vspace{\baselineskip}
\begin{justify}
To validate this theory, we first simulate a single cell type with a number of feature genes 3 times higher expressed than background genes on average(\textbf{b}). When the cells are divided into 2 clusters based on the PC1 (\textbf{c}), according to the Precision-Recall curve logistic regression shows certain but not ideal predictive ability in self-projection(\textbf{d}). Next, we simulate two cell types each defined by a set of feature genes that are non-overlapping(\textbf{e}). the PCA of these data are shown in (\textbf{f}). The self-projection shows perfect predictive ability(\textbf{g}). However, if the clustering tries to find three clusters, the cells of one of the types are over-clustered into 2 clusters(\textbf{h}). The over-clustered cell type shows lower accuracy(\textbf{i}). The confusion matrix(\textbf{j}) shows the size of overlaps between the original clusters and classes predicted by the trained model.The self-projection result demonstrates that confusion always happens between the over-clustered clusters, but not between clusters of different cell types. 
\end{justify}\par

\subsection*{Figure 2. Using the connection graph to optimise the clustering.}
\addcontentsline{toc}{subsection}{Figure 2. Using the connection graph to optimise the clustering.}
\begin{justify}
A simulated data (see \textbf{Methods}: multivariate normal simulation) set of six $``$putative cell types$"$ , the t-SNE plot of which is shown in \textbf{(a)}. The self-projection of the ideal cluster assignment is identical to the original clustering, with a self-projection accuracy $ \sim $ 94$\%$ . Starting from an over-clustered k-means clustering of 12 clusters, \textbf{(b)}, the confusion rates point to the clusters that represent the same putative cell type. The total accuracy is low in this starting point. To optimise the clustering, we first normalise the confusion matrix based on the number of correctly assigned cells in each cluster. Then, we remove the diagonal values and use a cutoff of normalized confusion rate to binarize the normalized confusion matrix into a \textit{connection matrix}, \textbf{(c)}. A \textit{connection graph}, \textbf{(d)}, is obtained from the connection matrix and is then used to merge some clusters. (Here we assume that the logistic regression model should achieve an accuracy better than 60$\%$  leaving a confusion rate threshold of 40$\%$ . If two cell clusters have a normalized confusion rate higher than 40$\%$  in the normalized confusion matrix, the two cell clusters are connected in the matrix.) SCCAF uses a Louvain clustering\href{https://paperpile.com/c/BKye4I/Jazbe}{\textsuperscript{12}} of fixed resolution of 1.0 to merge the identical cell clusters based on the binary connection graph. As shown in \textbf{(e)}, the merged clusters recover the initial cell type assignment well, except for some noisy cells in the center of the t-SNE plot, which are clustered as cluster 0. Finally, the self-projection of the logistic regression model optimizes the cell cluster assignment of the noisy cells. The optimized result was assessed by the SCCAF self-projection and attained accuracies of $ \sim $ 94$\%$  on both cross-validation and the test set, which is identical to the original simulated cell clusters \textbf{(a)}. The top weighted features (feature genes) captured by the logistic regression model are then compared with the feature genes that were used to simulate the cell clusters. The majority of feature genes are recapitulated by the logistic regression model \textbf{(f)}. 
\end{justify}\par

\subsection*{Figure 3. Self-projection-based clustering optimization compared with ground truth}
\addcontentsline{toc}{subsection}{Figure 3. Self-projection-based clustering optimization compared with ground truth}
\begin{justify}
In a mouse retina dataset (Shekhar et al. 2016), (\textbf{a}) Louvain clustering with resolution 0.3 and 1.0 result in over-clustering of the Rod Bipolar cells (red circle) and under-clusterings (blue circle). The SCCAF clustering optimization starts from a Louvain clustering of resolution 1.0. SCCAF merges the cell clusters based on the confusion matrix derived from the machine learning model. The confusion matrix-derived connection graph of the first-round optimization in (\textbf{b}) indicates that clusters 0,1,4,16 are connected. The optimization process increases the consistency between the clustering assignment and the model prediction. (\textbf{c}) Starting from louvain clustering resolution at 3.5, the cross-validation accuracies, and accuracies on the test sets increase over the four rounds of clustering optimization. The optimized result (\textbf{d}) is identical to an ideal annotation and is demonstrated by a river plot (\textbf{e}). 
\end{justify}\par

\subsection*{Figure 4. Self-projection accuracy indicates the optimal clustering during clustering optimization. }
\addcontentsline{toc}{subsection}{Figure 4. Self-projection accuracy indicates the optimal clustering during clustering optimization. }
\begin{justify}
In the datasets of Shekhar and Segerstolpe, the self-projection optimization result after four rounds are most similar to the gold standard annotation from the publication (\textbf{b} and \textbf{e}). Over-clusterings (\textbf{c} and \textbf{f}) exist in the results of the third round (red circles) and under-clusterings can be found in the fifth round (blue circles) results. The self-projection-based clustering optimization stops after four rounds, while further clustering merging can happen when we lower the confusion rate cutoff. According to the results (\textbf{a} and \textbf{d}) of 100 repeats random sampling of the self-projection, the optimal clustering demonstrates better self-projection accuracy than cases with over-clustering or under-clustering. 
\end{justify}\par

\subsection*{Figure 5. SCCAF captures key stages in mouse hematopoiesis. }
\addcontentsline{toc}{subsection}{Figure 5. SCCAF captures key stages in mouse hematopoiesis. }
\begin{justify}
Mouse hematopoiesis data from \href{https://paperpile.com/c/BKye4I/RdCh9}{\textsuperscript{41}} \textit{et al}. is clustered with SCCAF. The SCCAF clustering optimization starts from a Louvain clustering (\textbf{a}) and merged into 12 cell clusters (\textbf{b}). The resulting cell clusters are highly discriminative in the logistic regression model (\textbf{c}). The features encoded in the logistic regression model captures many of the known marker genes reported in previous publications (\textbf{d}). The top ranked features are listed, and known marker genes are highlighted in red. The cell clusters correspond to different cell types: Ba, basophilic or mast cell; D, dendritic; E, erythroid; GN, granulocytic neutrophil; Ly, lymphocytic; M, monocytic; Meg, megakaryocytic; MPP, multipotential progenitors; CEP, committed erythroid progenitors; ETD, erythroid terminal differentiation. Further, the erythrocytes are further clustered into 3 subpopulations, which can be clearly separated in the PCA spaces (\textbf{e}). These 3 subpopulations correspond to committed erythroid progenitors, early erythroids and late erythroids. Using the logistic regression model trained on the Tusi dataset and applied to another mouse hematopoiesis dataset from Giladi \textit{et al}. Most of the cell groups are recapitulated (\textbf{f}). The separation of the 3 erythroid stages of different biological functions (\textbf{h}) are well captured, while the self-projection accuracy is 90$\%$  on the Giladi dataset (\textbf{g}). 
\end{justify}\par

\printbibliography
\end{document}